\documentclass[aps,pra,pdf,superscriptaddress,twocolumn,showpacs,nofootinbib]{revtex4-2}
\usepackage[T1]{fontenc}
\usepackage{amsmath,amssymb,amsfonts}
\usepackage{eqnarray}
\usepackage{graphics,float}
\usepackage{bm}
\usepackage{braket}
\usepackage{amsmath}
\usepackage{physics}
\usepackage{mathtools}
\usepackage{gensymb}
\usepackage{color,xcolor,colortbl}
\usepackage{tikz}
\usepackage{soul}
\usetikzlibrary{shapes}
\usepackage[colorlinks=true,
linkcolor=blue,
filecolor=magenta,      
urlcolor=blue,
citecolor=blue]{hyperref}

\usepackage{lipsum}
\usepackage[normalem]{ulem} 
\usepackage{orcidlink}

\definecolor{redbrown}{rgb}{0.9882, 0.0196, 0.0196}

\newcommand{\markers}[2]{%
\raisebox{-0.2pt}{%
\tikz{
\node[
    star,
    star points=5,
    star point ratio=2.25,
    draw=#1,
    fill=#1,
    line width=0.7pt,
    minimum size=5pt,
    inner sep=0pt,
    #2
] {};
}%
}}

\graphicspath{{images/}}

\begin{document}
	\title{Mass-anisotropy driven stripe pattern formation and directional modulational instability in polariton condensate}
	\author{Hari Sadhan Ghosh\orcidlink{0009-0003-9969-390X}}
	\email{harisadhanghosh4@gmail.com}
	\affiliation{Department of Physics, Indian Institute of Technology Kharagpur, Kharagpur 721302, India}
	
	\author{Soumyadeep Halder\orcidlink{0000-0003-2824-8879}}
	\affiliation{ Indian Institute of Technology Gandhinagar, Palaj, Gujarat 382355, India}
    \affiliation{Max Planck Institute for the Physics of Complex Systems, Nöthnitzer Strasse 38, D-01187 Dresden, Germany}

	\author{Subrata Das\orcidlink{0000-0003-4527-8308}}
	\affiliation{Department of Physics, Virginia Tech, Blacksburg, Virginia 24061, USA}

	 \author{Sonjoy Majumder\orcidlink{0000-0001-9131-4520}}
	\email{sonjoym@phy.iitkgp.ac.in}
	\affiliation{Department of Physics, Indian Institute of Technology Kharagpur, Kharagpur 721302, India}
	
	\date{\today}
\begin{abstract}
Motivated by recent experimental advances toward electrical control of the effective mass and the realization of anisotropic mass distributions, we investigate the impact of mass anisotropy on driven-dissipative polariton condensates, focusing on modulational instability, pattern formation, and the collective dynamics of the emergent patterns. We show that unequal effective masses along the two orthogonal directions induce stripe pattern formation in an exciton-polariton condensate pumped by a non-resonant Gaussian laser beam above a critical mass ratio. A Bogoliubov analysis of the homogeneously pumped system reveals that mass anisotropy leaves the modulational-instability criteria unchanged but reshapes the unstable-mode dispersion by reducing the most unstable wavelength along the direction of larger effective mass with increasing mass ratio. This introduces an anisotropy-controlled length scale for pattern selection. When the pump width becomes comparable to the most unstable wavelength, the condensate becomes modulationally unstable and develops stripe patterns along the direction of larger effective mass, accompanied by particle-number fluctuations and center-of-mass oscillations along the same direction. As the pump width increases further, the stripes become distorted and develop spatial irregularities during the time evolution, indicative of modulational instability. Nevertheless, the anisotropic mass distribution preserves a pronounced directional preference, in contrast to the fully disordered patterns observed in the isotropic-mass case. Our results identify mass anisotropy as a key control parameter for directional modulational instability and pattern formation in nonequilibrium polariton condensates.
\end{abstract} 
	
\maketitle

\section{Introduction}
Exciton-polaritons are hybrid light-matter bosonic quasiparticles formed due to the strong coupling between quantum-well excitons and cavity photons in microcavities \cite{weisbuch_1992_observation,kavokin_2007_microcavities}. Due to its extremely low effective mass inherited from the photonic component and strong nonlinear interactions arising from the excitonic component, the exciton-polariton offers an excellent platform for exploring room-temperature Bose-Einstein condensation \cite{kasprzak_2006_bose,christopoulos_2007_room,kena_2010_room,plumhof_2014_room,deng_2002_condensation}. These systems have emerged as an ideal candidate to observe a wide range of novel phenomena, such as superfluidity \cite{amo_2009_superfluidity}, multistability \cite{lien_2015_multistability,ma_2018_vortex,yu_2021_non}, quantum hydrodynamics \cite{amo_2009_collective,nardin_2011_hydrodynamic,amo_2011_polariton,grosso_2011_soliton,saltykova_2025_quantum}, quantized vortices \cite{lagoudakis_2008_quantized,roumpos_2011_single,sanvitto_2011_all,nardin_2011_hydrodynamic,dominici_2018_interactions,gnusov_2023_quantum,gnusov_2024_vortex,keeling_2008_spontaneous,borgh_2012_robustness,sun_2019_emergence,kwon_2019_direct,yulin_2023_vorticity,gorbach_2010_vortex,gladilin_2019_noise,raman_2026_dynamically,fraser_2009_vortex,ma_2026_vortices,lagoudakis_2009_observation,lagoudakis_2011_probing,boulier_2015_vortex,panico_2021_dynamics,saito_2012_benard}, optical information processing \cite{ballarini_2013_all}, solitons and their dynamics \cite{zezyulin_2018_spin,xue_2014_creation,zhang_2021_generation,maitre_2020_dark,hu_2024_dark,hu_2025_gaussian,ostrovskaya_2012_dissipative,smirnov_2014_dynamics,hu_2026_self,egorov_2009_bright,sich_2012_observation,amo_2011_polariton,grosso_2011_soliton,grosso_2012_dynamics,yulin_2008_dark,kamchatnov_2012_oblique}, vortex molecules \cite{hu_2025_vortex}, turbulence \cite{ferrini_2025_driven,berloff_2010_turbulence,panico_2023_onset,comaron_2025_dynamics,koniakhin_2020_2d}, non-hermitian physics \cite{gao_2015_observation,yu_2021_non,dhara_2025_zero}, and Berezinskii-Kosterlitz-Thouless phase transition \cite{dagvadorj_2015_nonequilibrium,caputo_2018_topological,dagvadorj_2023_unconventional,comaron_2025_coherence}, among others.

In contrast to atomic condensates, exciton-polariton condensates are inherently non-equilibrium systems. Indeed, the stationary state does not correspond to the thermal equilibrium state, but rather originates from a dynamic balance between pumping and losses \cite{deng_2010_exciton,carusotto_2013_quantum}. This driven-dissipative feature gives rise to a rich variety of non-equilibrium dynamical phenomena, such as modulational instability \cite{bobrovska_2019_critical,bobrovska_2015_adiabatic,smirnov_2014_dynamics,bobrovska_2014_stability}, pattern formation \cite{whittaker_2017_polariton,alperin_2022_emergence,saito_2013_order,sanvitto_2006_spatial,ardizzone_2013_formation,li_2016_azimuthons}, chaotic dynamics \cite{bobrovska_2015_adiabatic,yaya_2023_routes,ruiz_2020_autonomous,gavrilov_2018_polariton}, etc., which have been extensively investigated both theoretically and experimentally. Among these phenomena, spatial pattern formation has attracted considerable attention in recent years owing to its close connection with the emergence of supersolid phases \cite{chomaz_2019_long-lived, chomaz_2023_dipolar, halder_2022_control,halder_2023_two-dimensional, ghosh_2024_induced, halder_2024_induced, halder_2025_roadmap, halder_2026_dynamical} in exciton-polariton condensates \cite{nigro_2025_supersolidity,trypogeorgos_2025_emerging,tian_2025_towards,zhai_2025_electrically,figueiredo_2026_supersolid, chen_2025_tunable, grudinina_2026_collective}. Various mechanisms have been shown to generate patterns, including nonlinear interactions \cite{wouters_2007_excitations}, spin-orbit coupling \cite{borgh_2010_spatial,muszynski_2024_observation, saboo_2025_magnetization}, external potentials \cite{dong_2024_dynamical,berman_2015_spontaneous}, and pump engineering \cite{manni_2011_spontaneous,saito_2013_order}.

More recently, electrical control of the effective mass of polariton has emerged as a powerful tool for manipulating vortex charge in the condensate \cite{zhai_2023_electrically}. Motivated by these developments, we investigate the modulationally unstable behavior and the resulting stripe-pattern formation in a non-resonantly pumped exciton-polariton condensate with anisotropic effective masses along two orthogonal directions. We show that, above a critical mass ratio, the mass anisotropy gives rise to stripe pattern-forming dynamics. The stripe pattern and self-oscillation of the system along the direction of the greater effective mass arise spontaneously as a result of the anisotropic mass distribution breaking the translational symmetry. This self-oscillation allows the system to acquire linear momentum in the direction of the greater effective mass, resulting in oscillating current dynamics in that direction. Our findings open an experimentally feasible route to engineer stripe pattern formation and collective dynamics in a driven-dissipative polariton condensate.

The structure of this paper is as follows. In Sec. \ref{formalism}, we introduce the theoretical model for the description of polariton condensate. In Sec. \ref{homogeneous}, we perform the linear stability analysis of the homogeneous steady state and determine the instability spectrum. In Sec. \ref{inhmogeneous} we investigate the stability and instability regions under an inhomogeneous pumping profile for different system parameters and examine the resulting stripe-pattern formation and self-oscillatory dynamics. Finally, in Sec. \ref{conclusion} we summarize our findings.
 
 \section{Formalism}
\label{formalism}
 Within the mean-field description, the dynamics of the exciton-polariton condensate with macroscopic wave function $\psi_c(\mathbf{r},t)$ is modeled by an open-dissipative Gross-Pitaevskii equation coupled to the rate equation for the density of the polariton reservoir $n_r(\mathbf{r},t)$ \cite{wouters_2007_excitations,carusotto_2013_quantum,wouters_2009_stochastic}.
 \begin{align}
    i\hbar\partial_t\psi_c&(\mathbf{r},t)= \Bigg[-\frac{\hbar^2\Delta_x}{2m_x} -\frac{\hbar^2\Delta_y}{2m_y} + g_c|\psi_c(\mathbf{r},t)|^2 \nonumber  \\
    & -i\hbar\frac{\gamma_c}{2}+\left(g_r+i\hbar\frac{R}{2}\right)n_r(\mathbf{r},t)\Bigg]\psi_c(\mathbf{r},t),
    \label{cgpe}
 \end{align} 
 \begin{equation}
    \partial_t n_r(\mathbf{r},t)=P(\mathbf{r})+\left[-\gamma_r-R|\psi_c(\mathbf{r},t)|^2\right]n_r(\mathbf{r},t),
    \label{rpol}
 \end{equation}
 where $\Delta_{x(y)}$ and $m_{x(y)}$ are the Laplacian operators and the effective masses of the polariton condensates along the $x(y)$ directions. The effective mass is electrically tuned by changing the direction of orientation of the liquid crystal molecules \cite{zhai_2023_electrically}. The condensate decays at a rate $\gamma_c$ and is replenished through stimulated scattering from the reservoir at a condensation rate $R$. $g_c$ represents the self-interaction between the condensate polaritons, and $g_r$ is the interaction constant between the reservoir polaritons and the condensate polaritons. Here, polaritons are injected non-resonantly into the reservoir through the pump beam profile $P(\mathbf{r})$, while the reservoir population decays at a rate $\gamma_r$.

 \section{Effects of the mass-anisotropy on the stability of a homogeneous condensate}
 \label{homogeneous}
 Before analyzing the impact of anisotropic mass on the steady-state solution of the polariton condensate under spatially uniform pumping, $P(\mathbf{r})=P$, we discuss the uniform steady-state solution $\psi_c(\mathbf{r},t)=\psi_0e^{-i\mu_0 t/\hbar}$ and $n_r(\mathbf{r},t)=n_r^0$ with homogeneous pumping $P$, and with isotropic mass distribution $m_x=m_y$. As derived from Eqs. (\ref{cgpe}) and (\ref{rpol}), for $P$ being less than the threshold pump power $P_{\rm{th}}=\gamma_c\gamma_r/R$, there is no formation of condensate $|\psi_0|^2=n_c^0=0$; on the other hand, the density of the reservoir is proportional to the intensity of the pump as $n_r^0=P/\gamma_r$. Above the threshold $P>P_{\rm{th}}$, a finite condensate density $|\psi_0|^2=n_c^0=(P-P_{\rm{th}})/\gamma_c$ emerges,  and the reservoir density saturates at $n_r^0=\gamma_c/R$ with chemical potential $\mu_0=g_c|\psi_0|^2+g_rn_r^0$ \cite{carusotto_2013_quantum,wouters_2007_excitations}.
 
 In the following, we explore how deviations from mass isotropy modify the steady state properties, particularly its density profile and stability. As the anisotropic mass distribution $m_x\neq m_y$ modifies only the kinetic energy operator $-\hbar^2\Delta_{x(y)}/2m_{x(y)}$, the steady-state solution under homogeneous pumping is unaffected. However, it remains to be determined whether this mass anisotropy alters the conditions for modulational stability of a homogeneous background. We therefore employ the Bogoliubov-de Gennes approximation \cite{pitaevskii_2016_bose,pethick_2008_bose} to perform a linear stability analysis, assuming a small perturbation around the steady state solutions \cite{carusotto_2013_quantum,wouters_2007_excitations,byrnes_2012_negative}
 \begin{equation}
\psi_c(\mathbf{r},t) = e^{-i\mu_0 t/\hbar}\Big[\psi_0 + \delta\psi_c(\mathbf{r},t)\Big],
\label{psi}
\end{equation}
\begin{equation}
n_r(\mathbf{r},t) = n_r^0 + \sum_\mathbf{k} \left[ c_\mathbf{k} e^{-i\omega_\mathbf{k} t + i \mathbf{k}.\mathbf{ r}} + c_\mathbf{k}^{*} e^{i\omega_\mathbf{k}^{*} t - i \mathbf{k}.\mathbf{ r}} \right],
\label{nr}
\end{equation}
where $\delta\psi_c(\mathbf{r},t)=\sum_\mathbf{k} ( a_\mathbf{k} e^{-i\omega_\mathbf{k} t + i \mathbf{k}.\mathbf{ r}} + b_\mathbf{k} e^{i\omega_\mathbf{k}^{*} t - i \mathbf{k}.\mathbf{ r}})$ is the small fluctuation around the steady state, $\omega_\mathbf{k}$ is the excitation frequency of the mode with wave vector $\mathbf{k}$, and $a_\mathbf{k}$, $b_\mathbf{k}$, $c_\mathbf{k}$ are the corresponding mode coefficients. By substituting Eqs. (\ref{psi}) and (\ref{nr}) into Eqs. (\ref{cgpe}) and (\ref{rpol}) and linearizing them, we obtain the eigenvalue problem $\mathcal{L}_{\mathbf{k}}\mathcal{M}_{\mathbf{k}}=\hbar\omega_{\mathbf{k}}\mathcal{M}_{\mathbf{k}}$ for the elementary excitations with $\mathcal{M}_{\mathbf{k}}=(a_\mathbf{k}$,$b_\mathbf{k}$,$c_\mathbf{k})^T$ and

\begin{equation}
\mathcal{L}_{\mathbf{k}} =
\begin{pmatrix}
\epsilon_\mathbf{k}+g_cn_c^0 &
g_cn_c^0 &
(i\hbar \frac{R}{2}+g_r)\sqrt{n_c^0} \\[3pt]
-g_cn_c^0 &
-\epsilon_{\mathbf{k}}-g_cn_c^0 &
 (i\hbar \frac{R}{2}-g_r)\sqrt{n_c^0}\\[3pt]
-i\hbar\gamma_c \sqrt{n_c^0}&
 -i\hbar\gamma_c \sqrt{n_c^0}&
-i\hbar\gamma_r-i\hbar R n_c^0
\end{pmatrix},
\label{eq:Lk}
\end{equation}
with $\epsilon_{\mathbf{k}}=\hbar^2(k_x^2/2m_x + k_y^2/2m_y)$ being the kinetic energy. The dispersion relation of the system is obtained by solving the eigenvalue problem, resulting in a cubic equation with respect to $\omega_{\mathbf{k}}$ as follows
\begin{equation}
\begin{split}
    \hbar^3\omega_\mathbf{k}^3 + (i\hbar\gamma_r + i\hbar R n_c^0)\hbar^2\omega_{\mathbf{k}}^2 -(\omega_B^2 + \hbar^2 R \gamma_c n_c^0)\hbar\omega_{\mathbf{k}} \\ =
    (i\hbar\gamma_r +i\hbar R n_c^0)\omega_B^2 -2i\hbar\gamma_cg_r n_c^0\epsilon_{\mathbf{k}},
    \end{split}
    \label{dispersion}
\end{equation}
where $\omega_B^2=\epsilon_{\mathbf{k}}(\epsilon_{\mathbf{k}}+2g_c n_c^0)$. From Eq. \eqref{dispersion}, it is obtained in the regime $P_{\rm{th}}<P<\frac{g_r\gamma_c}{g_c\gamma_r}P_{\rm{th}}$ under the condition $\frac{g_r\gamma_c}{g_c\gamma_r}>1$, the imaginary part of the eigenvalue $\omega_{\mathbf{k}}$ becomes positive within finite intervals $k_{x(y)}\in\big[k_{x(y)}^{\rm min},k_{x(y)}^{\rm max}\big]$. Therefore, homogeneous condensate is dynamically unstable in this momentum range, where density modulations grow exponentially in time \cite{bobrovska_2019_critical,saito_2016_selfrotation,wouters_2007_excitations,bobrovska_2018_dynamical,estrecho_2018_single,baboux_2018_unstable,bobrovska_2014_stability,smirnov_2014_dynamics}. The dispersion relation, Eq. \eqref{dispersion}, yields the lower and upper momentum limits of the instability as follows:
\begin{align}
    &k_{x}^{\rm min}=k_{y}^{\rm min}=0,\\
    &k_{x(y)}^{\rm max}=\left[4g_cn_c^0m_{x(y)}\left(\frac{g_r\gamma_c}{g_c\gamma_r}.\frac{ P_{\rm th}}{P}-1\right)\right]^{1/2}.
     \label{limit}
\end{align}

\begin{figure}
	\centering
	\includegraphics[width=0.48\textwidth]{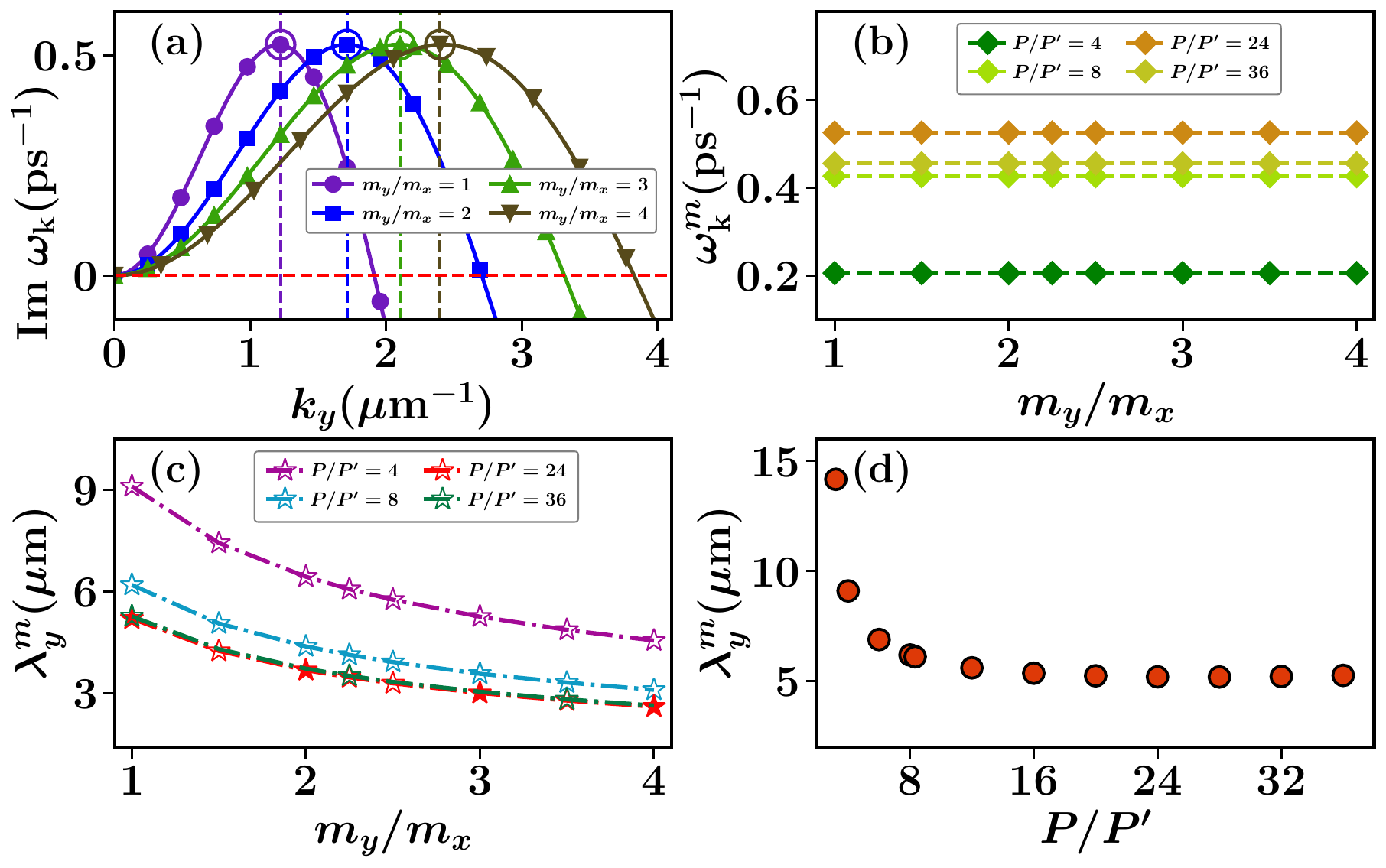}
	\caption{(a) Imaginary part of the selected Bogoliubov spectrum, $\mathrm{Im}~\omega_{\mathbf{k}}$, which has both positive and negative values, at pump power $P=24P^\prime$ with different mass ratio $m_y/m_x$. The circular markers in (a) mark the analytically calculated point $(k^m_y,\omega^m_{\mathbf{k}})$, while the dashed vertical lines indicate the corresponding values extracted numerically from Figs. \ref{fig:2}(b1)-\ref{fig:2}(b2). (b) Variation of $\omega^m_{\mathbf{k}}=\mathrm{Max}(\mathrm{Im}~\omega_{\mathbf{k}})$ with mass ratio at different pump strengths. (c) Variation of the most unstable wavelength $\lambda^m_y$ with mass ratio $m_y/m_x$ at different pump power $P$. The filled star markers (\protect\markers{redbrown}{}) denote the most unstable wavelength $\lambda_y^m$ for mass ratio $m_y/m_x=1$, $2$, $3$, and $4$. (d) shows the change in $\lambda^m_y$ with $P/P^\prime$ at fixed $m_y/m_x=1$. The parameters are $m_x=5\times10^{-5}m_e, m_e=$ electron mass, $g_c=6\times 10^{-3} ~\mathrm{meV}\mu\mathrm{m}^2, g_r=2g_c, R=0.04~\mathrm{ps}^{-1}\mu\mathrm{m}^2, \gamma_c^{-1}=0.15~\mathrm{ps}, \gamma_r^{-1}=3~\mathrm{ps}, P=460.6776~\mathrm{ps}^{-1}\mu\mathrm{m}^{-2}=24P^\prime, P^\prime=19.1949 ~\mathrm{ps}^{-1}\mu\mathrm{m}^{-2}$ \cite{saito_2016_selfrotation}.}
    \label{fig:1}
\end{figure}

So, our theoretical calculation confirms that the anisotropic mass distribution does not alter the instability condition of the homogeneous state, i.e., $P_{\rm{th}}<P<\frac{g_r\gamma_c}{g_c\gamma_r}P_{\rm{th}}$. Instead, it modifies the upper bounds $k_{x(y)}^{\rm max}$ of the unstable $k_{x(y)}$ intervals and shifts the most unstable wave vectors $k_x^{m}$ and $k_y^{m}$ along the corresponding directions [see Eq. \eqref{limit}]. Here, $k_x^m$ and $k_y^m$ denote the wave vectors at which the imaginary part of the eigenfrequency, $\mathrm{Im}~\omega_{\mathbf{k}}$, reaches its maximum. Fig. \ref{fig:1}(a) illustrates the change in $k_y^{m}$ with mass ratio $m_y/m_x$ for pump power $P=24P^\prime$ $\approx 8.29P_{\rm{th}}$. As $m_y/m_x$ increases, $k_y^m$ increases [see Fig. \ref{fig:1}(a)], and the corresponding most unstable wavelength $\lambda^m_y=2\pi/k_y^{m}$ decreases accordingly, as shown in Fig. \ref{fig:1}(c). But as $m_x$ is held fixed, the most unstable wavelength $\lambda_x^m=2\pi/k_x^m$ remains constant. In Figs. \ref{fig:1}(b) and \ref{fig:1}(c), we plot the variation of $\omega^m_{\mathbf{k}}=\mathrm{Max}(\mathrm{Im}~\omega_{\mathbf{k}})$ and $\lambda^m_y$ with the pump power $P$ and the mass ratio $m_y/m_x$, respectively. Notably, $\omega^m_{\mathbf{k}}$ differs at different pump powers $P$ and is independent of the mass ratio $m_y/m_x$.  On the other hand, the value of $\lambda^m_y$ alters with the mass ratio $m_y/m_x$ and the pump power $P$. However, with a higher pump power, its value becomes constant for a fixed mass ratio, as shown in Fig. \ref{fig:1}(d) for the isotropic case $m_y/m_x=1$.

\begin{figure}
	\centering
	
    \includegraphics[width=0.48\textwidth]{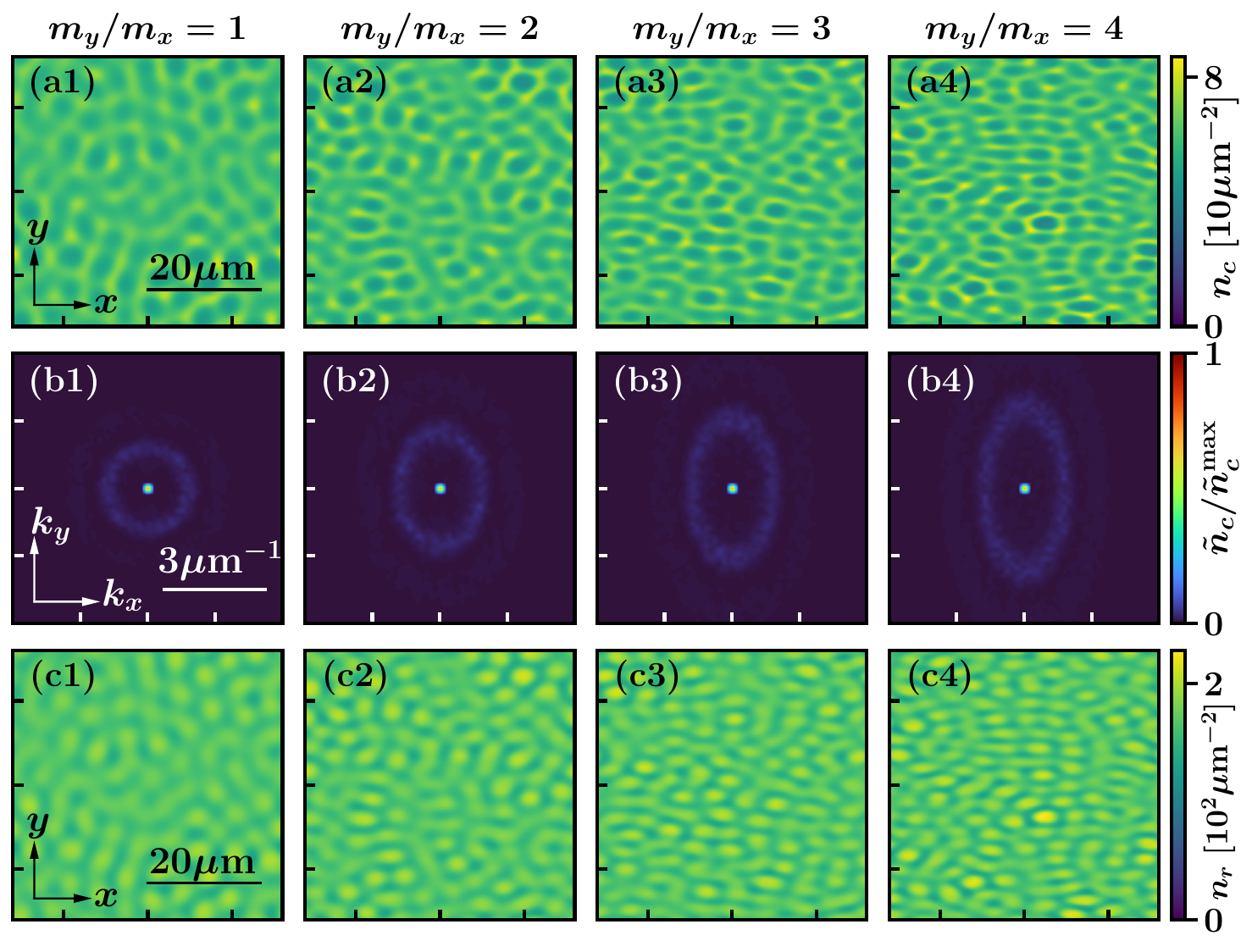}

	\caption{Pattern formation due to modulational instability. The upper panel [(a1)-(a4)] shows the modulated condensate density $n_c(\mathbf{r},t)=|\psi_c(\mathbf{r},t)|^2$ for different $m_y/m_x$ at $t=20.5 ~\mathrm{ps}$, while the middle panel [(b1)-(b4)] shows the corresponding momentum-space density $\tilde{n}_c(\mathbf{k},t)$. The lower panel [(c1)-(c4)] shows the reservoir densities $n_r(\mathbf{r},t)$ for same values of $m_y/m_x$. Simulation parameters are the same as in Fig. \ref{fig:1}.}
    \label{fig:2}
\end{figure}
To validate our analytical calculation, we numerically integrate Eqs. (\ref{cgpe}) and (\ref{rpol}) using the split-step Fourier method under spatially homogeneous pumping. The initial state is taken as $\psi_c(\mathbf{r},t=0)=\sqrt{n_c^0}$ and $n_r(\mathbf{r},t=0)=\gamma_c/R$, with small random noise added to seed the instability \cite{saito_2016_selfrotation}. We also make sure that the chosen pump power $P=24P^\prime\approx8.29P_{\rm{th}}$ lies between the instability condition $P_{\rm{th}}<P<\frac{g_r\gamma_c}{g_c\gamma_r}P_{\rm{th}}$ with $\frac{g_r\gamma_c}{g_c\gamma_r}=40$. To investigate the pattern formation due to modulational instability, we plot the condensate and reservoir densities in the upper and lower panels of Fig. \ref{fig:2}, respectively. And, it is clear that the spatially modulated high-density pattern is formed at $t=20.5~\rm{ps}$. Due to the repulsive interaction $g_r$ between the reservoir and the condensate polariton, high-density condensate regions form in the areas of low reservoir density, and vice versa. As $m_y/m_x$ increases, the number of high-density peaks increases both in the condensate and in the reservoir density along the greater mass direction. To identify the most unstable wave vectors, we perform a Fourier transformation of the condensate density $n_c(\mathbf{r},t)$ and plot the momentum-space density $\tilde{n}_c(\mathbf{k},t)$ in the middle row of Fig. \ref{fig:2}. Fig. \ref{fig:2}(b1) shows that, in addition to the central peak at $(k_x,k_y)=(0,0) ~\mu\mathrm{m}^{-1}$, a circular ring of radius $1.21~\mu\mathrm{m}^{-1}$ appears in the $k_x$-$k_y$ plane, indicating the density pattern in the real space. Now, as $m_y/m_x$ increases, the circular ring deforms to an elliptical one with the major axis along the higher mass direction. The length of the semi-minor and the semi-major axes of the ellipse represent the most unstable wave vectors $k_x^{m}$ and $k_y^{m}$ along the corresponding directions, respectively. The larger $k_y^{m}$ signifies the higher concentration peaks along the $y$-direction. For different mass ratios $m_y/m_x=1,2,3,$ and $4$, Figs. \ref{fig:2}(b1)-\ref{fig:2}(b4) give $(k_x^{m},k_y^{m})$: $(1.21,1.21)~\mu\mathrm{m}^{-1}, (1.21,1.71)~\mu\mathrm{m}^{-1}, (1.21,2.09)~\mu\mathrm{m}^{-1}$, and $(1.21,2.42)~\mu\mathrm{m}^{-1}$, respectively, which agree well with our analytical predictions [see Fig. \ref{fig:1}(a). The circular markers in Fig. \ref{fig:1}(a) mark the analytically calculated point $(k^m_y,\omega^m_{\mathbf{k}})$, while the dashed vertical lines indicate the corresponding values extracted numerically from Fig. \ref{fig:2}(b1)-\ref{fig:2}(b2)]. So, our calculations confirm $k_y^m/k_x^m$ is proportional to $\sqrt{m_y/m_x}$. By symmetry, for $m_y/m_x<1$ the roles reverse. Thus, in general, the ratio of the major to minor axis of the ellipse scales as $\sqrt{\max(m_y/m_x,m_x/m_y)}$, with the ellipse always elongated along the direction of larger effective mass.
\begin{figure}
	\centering
	\includegraphics[width=0.48\textwidth]{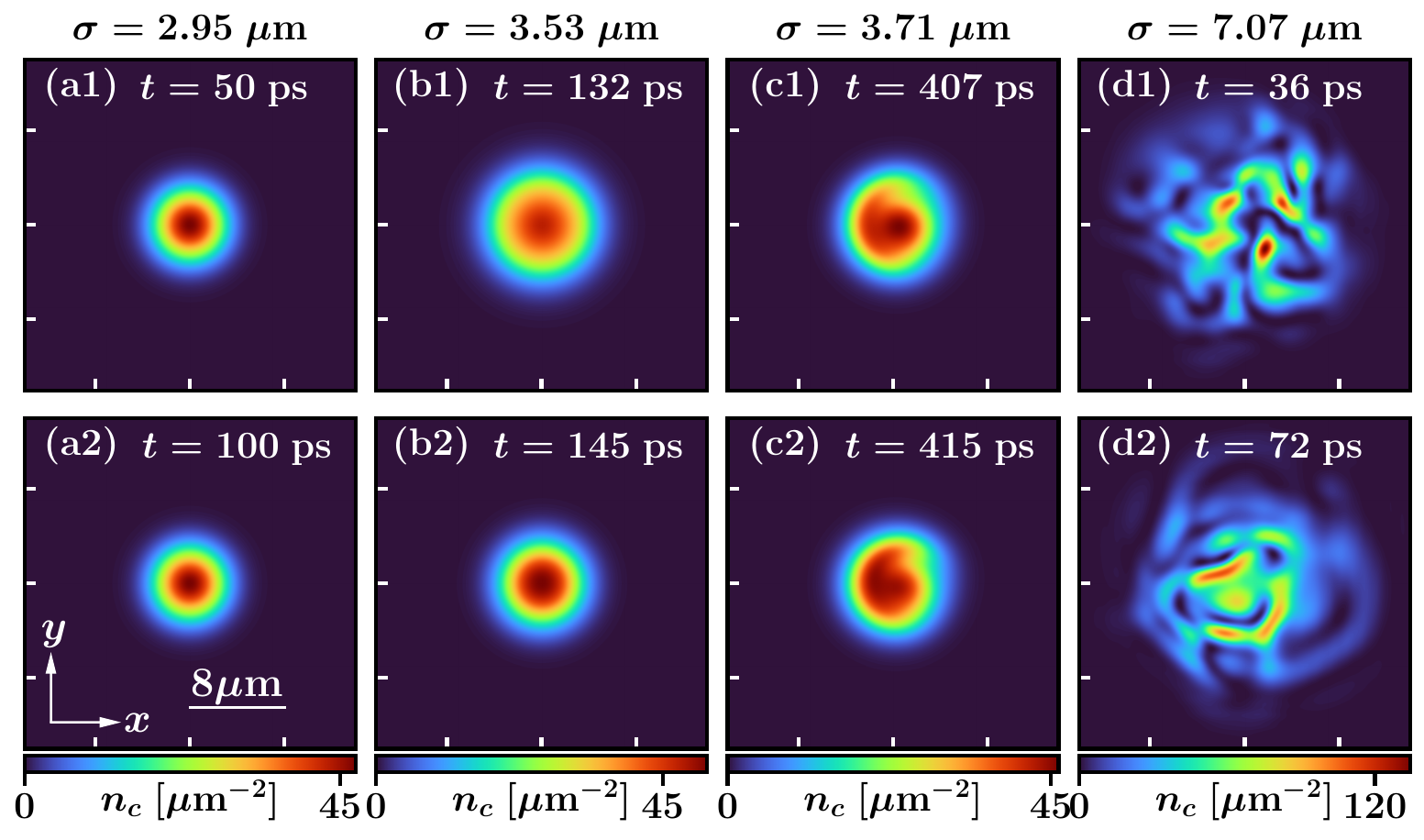}
	\caption{Time evolution of non-resonantly pumped polariton condensates under Gaussian pumping with different beam widths $\sigma$ and mass ratio $m_y/m_x=1$. (a1) and (a2) show stable Gaussian shaped condensate for beam width $\sigma=2.95 ~\mu \mathrm{m}$ at two different time instants. (b1) and (b2) display a breathing Gaussian shaped condensate for beam width $\sigma=3.53 ~\mu \mathrm{m}$. (c1) and (c2) are self-rotating condensates at two different time instants for beam width $\sigma=3.71 ~\mu \mathrm{m}$. (d1) and (d2) correspond to a modulationally unstable condensate exhibiting pattern formation for beam width $\sigma=7.07 ~\mu \mathrm{m}$. Other parameters are identical to those used in Fig. \ref{fig:1}.}
    \label{fig:3}
\end{figure}
 \section{Effects of the mass-anisotropy on the stability of an inhomogeneous condensate}	
 \label{inhmogeneous}
Having analyzed the homogeneous case of pump, we now extend our study to a polariton condensate with an anisotropic effective mass distribution under non-resonant pumping by the Gaussian-shaped laser beam with profile
\begin{equation}
P(\mathbf{r})=P\exp(-r^2/\sigma^2). 
\label{pump}
\end{equation}
Here, $P=24P^\prime\approx8.29P_{\rm{th}}$ is the maximum intensity and $\sigma$ is the $1/e$ width of the pump profile. Before addressing the anisotropic-mass system, we first revisit how the pump width $\sigma$ governs the stability of the condensate in the isotropic mass case. To perform the numerical simulations, both the condensate wave function $\psi_c(\mathbf{r},0)$ and the reservoir density $n_r(\mathbf{r},0)$ are initialized with small random noise. Other parameters are the same as in Fig. \ref{fig:1}.  
\begin{figure}
	\centering
	\includegraphics[width=0.46\textwidth]{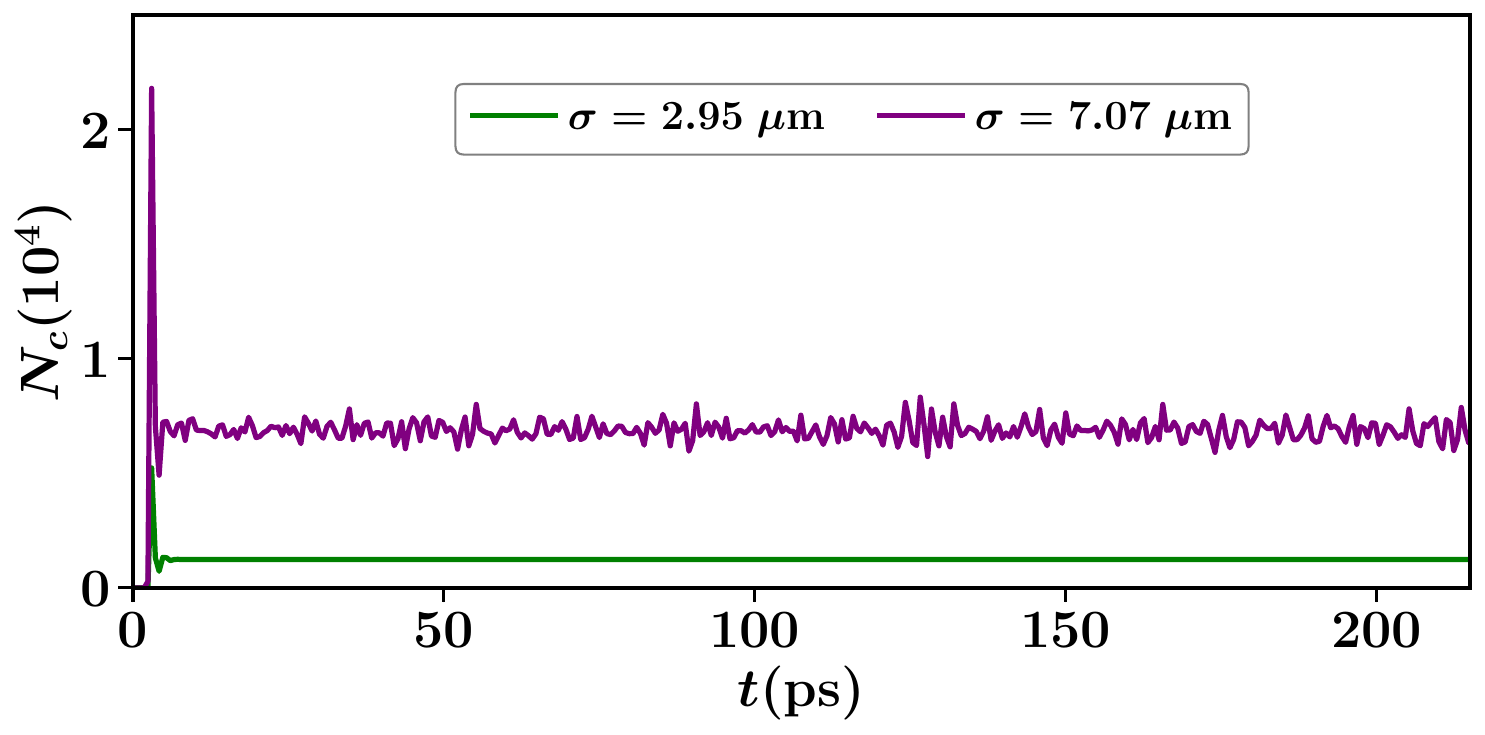}
	\caption{Time evolution of the condensate particle number $N_c=\int d\mathbf{r}~|\psi_c(\mathbf{r},t)|^2$ for beam widths $\sigma=2.95~\mu \mathrm{m}$ and $\sigma=7.07~\mu \mathrm{m}$ with mass ratio $m_y/m_x=1$ .}
    \label{fig:4}
\end{figure}

For a homogeneous condensate driven by a uniform pump $P(\mathbf{r})=P=24P^\prime\approx8.29P_{\rm th}$, the most unstable wavelengths for an isotropic mass distribution are $\lambda_x^{m}=\lambda_y^{m}=5.20~\mu\mathrm{m}$. We first consider a situation in which the Gaussian pump width, $\sigma=2.95~\mu\mathrm{m}$, is significantly smaller than these unstable wavelengths. Under this condition, the condensate remains modulationally stable and evolves into a stationary Gaussian-shaped state as shown in Figs. \ref{fig:3} (a1) and \ref{fig:3}(a2) \cite{saito_2016_selfrotation,bobrovska_2014_stability,baboux_2018_unstable,bobrovska_2018_dynamical}.
Increasing the width of the pump $\sigma$ to $3.33 ~\mu \mathrm{m}$ still yields a stable Gaussian condensate. When $\sigma$ becomes comparable to the most unstable wavelength, the condensate develops dynamical states such as self-breathing \cite{bobrovska_2014_stability} and self-rotation \cite{saito_2016_selfrotation,berman_2015_spontaneous}, as illustrated in Figs. \ref{fig:3}(b1), \ref{fig:3}(b2) and Figs. \ref{fig:3}(c1), \ref{fig:3}(c2) for $\sigma=3.53 ~\mu \mathrm{m}$ and $\sigma=3.71 ~\mu \mathrm{m}$, respectively. Upon further increasing $\sigma$, the condensate enters a modulationally unstable regime \cite{bobrovska_2014_stability,bobrovska_2018_dynamical}, leading to pattern formation, as shown in Figs. \ref{fig:3}(d1) and \ref{fig:3}(d2).
In addition, to characterize the stability of the condensate, we plot the time evolution of the number of condensate particles for $\sigma=2.95 ~\mu \mathrm{m}$ and $\sigma=7.07 ~\mu \mathrm{m}$ in Fig. \ref{fig:4}. We find that for a modulationally stable polariton condensate $[\sigma\ll (\lambda_x^m,\lambda_y^m)]$, the number of condensate polaritons increases from zero and eventually saturates at a constant value determined by the Gaussian pump width $\sigma$. In contrast, for larger pump widths $\sigma> \lambda_x^m$ or $\sigma>\lambda_y^m$, pronounced fluctuations in the condensate population are observed throughout the time evolution [see Fig. \ref{fig:4}], providing a clear signature of modulational instability.

Now, we are going to explore the influence of mass-anisotropy on the polariton condensate pumped by a Gaussian-laser beam by varying the mass ratio $m_y/m_x$ from 1 to 4 for different pump widths $\sigma$. We introduce mass anisotropy by increasing the effective mass along the $y$-direction while keeping the effective mass along the $x$-direction fixed \cite{zhai_2023_electrically}. Consequently, the most unstable wavelength along the $x$-direction remains unchanged, whereas $\lambda_y^m$ decreases with increasing mass ratio $m_y/m_x$ [see Eq. \eqref{limit}]. The analytically evaluated most unstable wavelengths of the homogeneous condensate $(\lambda_x^m,\lambda_y^m)$ for $m_y/m_x=1,2,3,$ and $4$ are $(5.20,5.20)~\mu\mathrm{m}, (5.20,3.68)~\mu\mathrm{m}, (5.20,3.00)~\mu\mathrm{m}$ and $(5.20,2.60)~\mu\mathrm{m}$, respectively. This reduction of the most unstable wavelength with increasing mass anisotropy plays a crucial role in the emergence of directional modulational instability.
\begin{figure}
	\centering
	\includegraphics[width=0.48\textwidth]{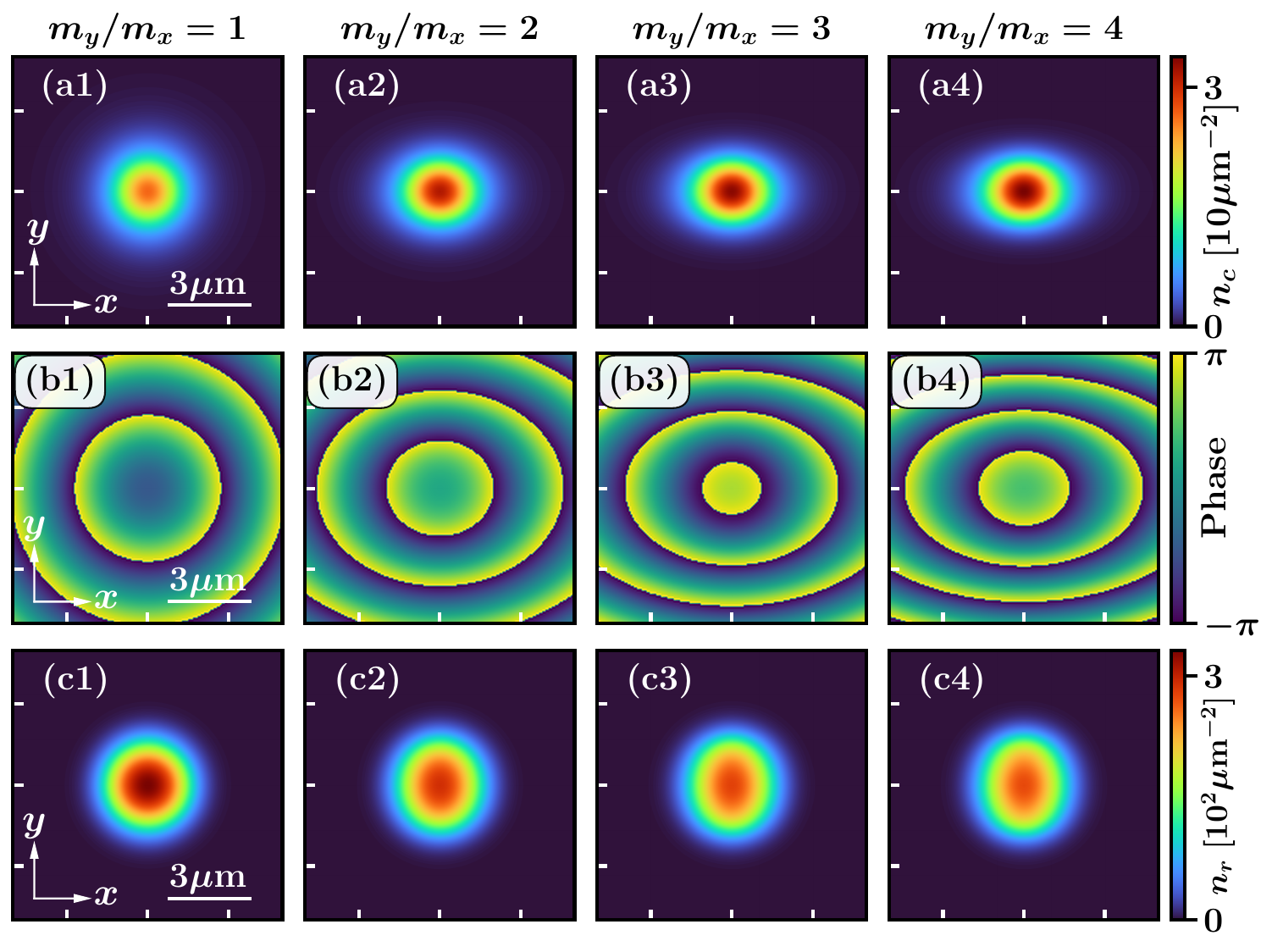}
	\caption{The upper panel [(a1)-(a4)] shows the density $n_c(\mathbf{r},t)=|\psi_c(\mathbf{r},t)|^2$ of condensate for different mass ratios $m_y/m_x$ with pump width $\sigma=1.18~\mu\mathrm{m}$. The corresponding condensate phase profiles are displayed in the middle panel [(b1)-(b4)], while the associated reservoir densities $n_r(\mathbf{r},t)$ are shown in the lower panel [(c1)-(c4)]. Other parameters are the same as in Fig. \ref{fig:1}.}
    \label{fig:5}
\end{figure}

\begin{figure}
	\centering
	\includegraphics[width=0.46\textwidth]{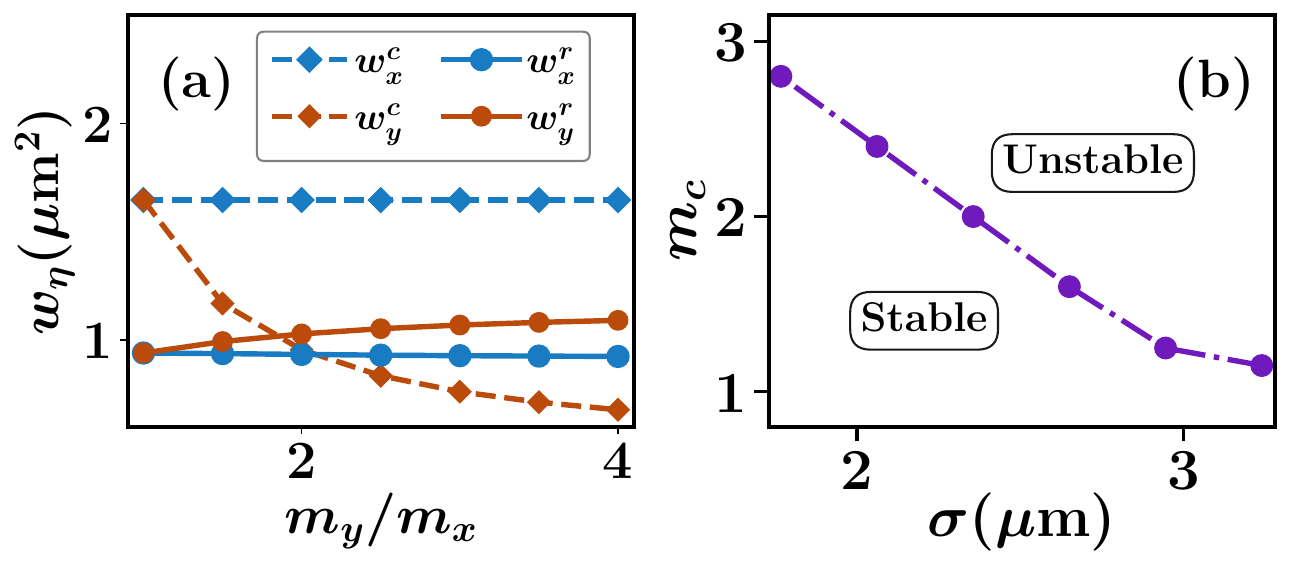}
	\caption{(a) Condensate and reservoir mean-square widths $w_x^{c(r)}$ and $w_y^{c(r)}$, after the system reaches the stationary state, as a function of mass ratios $m_y/m_x$ for beam width $\sigma=1.18~\mu \mathrm{m}$. (b) Variation of critical mass ratio $m_c$ with beam width $\sigma$, above which the condensate becomes unstable. Other parameters are the same as in Fig. \ref{fig:1}.}
    \label{fig:6}
\end{figure}

First, we focus on Gaussian pump widths for which the mass-isotropic system is modulationally stable, i.e., $\sigma \ll \lambda_y^m$. For $\sigma=1.18~\mu\mathrm{m}$, the mass-anisotropic system remains modulationally stable up to $m_y/m_x=4$, as the pump width is much smaller than the corresponding most unstable wavelengths, $(\lambda_x^m,\lambda_y^m)=(5.20,2.60)~\mu\mathrm{m}$. Although the system eventually reaches a steady state and remains modulationally stable, increasing the mass anisotropy causes the condensate profile to deform from a symmetric Gaussian into an elliptical shape with major axis along the $x$-direction, as shown in the upper panel of Figs. \ref{fig:5}(a1)-\ref{fig:5}(a4). With increasing mass ratio $m_y /m_x$, the mean-square width of the condensate $w_y^c=\int d\mathbf{r}~y^2\abs{\psi_c(\mathbf{r},t)}^2/N_c$ along the $y$-direction decreases, while the width $w_x^c=\int d\mathbf{r}~x^2\abs{\psi_c(\mathbf{r},t)}^2/N_c$ along the $x$-direction remains unchanged, as shown in Fig. \ref{fig:6}(a).
Simultaneously, the condensate phase profiles change from a circularly symmetric structure to an elliptical one with increasing mass anisotropy, as shown in the middle panel of Fig. \ref{fig:5}. As the shape of the condensate changes, it also modifies the density profile of the reservoir polariton. Owing to the repulsive interaction between the condensate and reservoir polaritons, the reservoir density also becomes elliptical, with its major axis oriented along the $y$-direction, as shown in Figs.~\ref{fig:5}(c1)-\ref{fig:5}(c4). To identify the change in the reservoir density profile we plot its mean-square widths in Fig.~\ref{fig:6}(a). In contrast to the condensate, the mean-square width of reservoir $w_y^r=\int d\mathbf{r}~y^2 n_r(\mathbf{r},t)/\int d\mathbf{r}~ n_r(\mathbf{r},t)$ increases with the mass ratio $m_y/m_x$, while $w_x^r=\int d\mathbf{r}~x^2 n_r(\mathbf{r},t)/\int d\mathbf{r}~ n_r(\mathbf{r},t)$ remains unchanged.

\begin{figure*}[t]
	\centering
	\includegraphics[width=\textwidth]{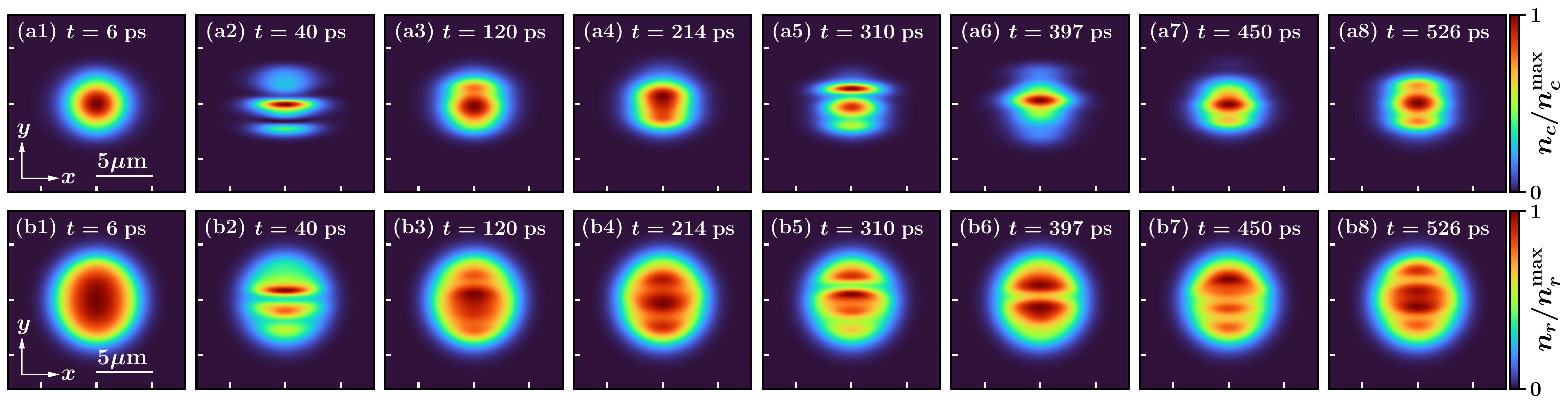}
	\caption{Stripe-pattern formation due to directional modulational instability. (a1)-(a8) show the condensate density $n_c(\mathbf{r},t)=|\psi_c(\mathbf{r},t)|^2$ at different time instants for mass ratio $m_y/m_x=4$ and beam width $\sigma=2.36~\mu\mathrm{m}$. (b1)-(b8) show the corresponding reservoir density profiles $n_r(\mathbf{r},t)$. Other parameters are the same as in Fig. \ref{fig:1}. See the Supplemental Material for a movie of the dynamics.}
    \label{fig:7}
\end{figure*}

\begin{figure}[b]
	\centering
	\includegraphics[width=0.48\textwidth]{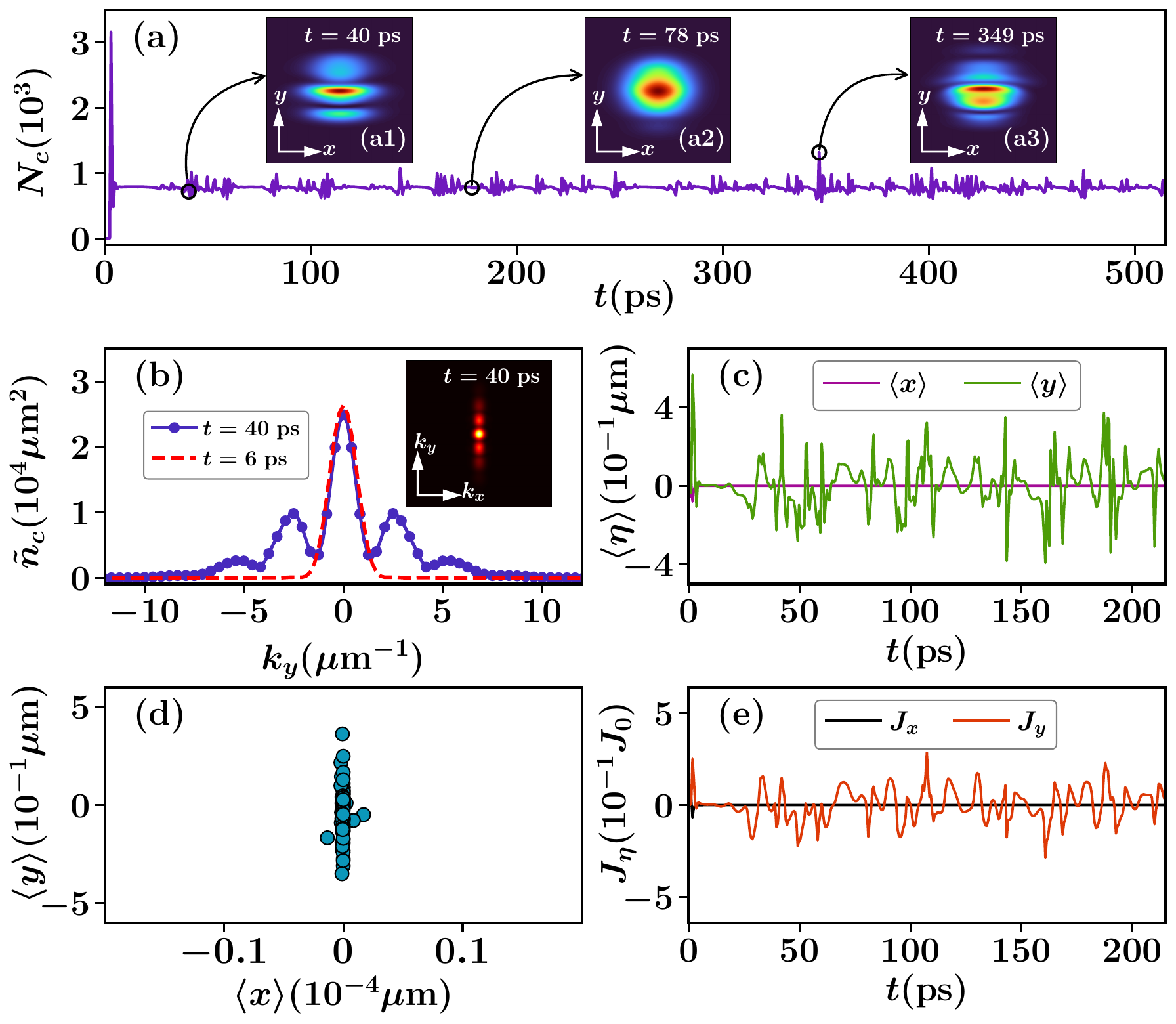}
	\caption{(a) Time evolution of the condensate particle number $N_c$ for beam width $\sigma=2.36~\mu \mathrm{m}$ and mass ratio $m_y/m_x=4$. Insets (a1), (a2) and (a3) shows the density distribution at $t=40~\rm{ps}$, $t=78~\rm{ps}$, and $t=349~\rm{ps}$ respectively. At $t=78~\rm{ps}$, the condensate exhibits an unmodulated density distribution with negligible particle-number fluctuations. But, at $t=40~\rm{ps}$ and $t=349~\rm{ps}$, significant particle-number fluctuations emerge, accompanied by the formation of a spatially modulated density pattern. (b) The momentum-space density $\tilde{n}_c(k_x=0,k_y)$ at $t=6~\mathrm{ps}$ and $t=40~\mathrm{ps}$. The inset shows the momentum-space density distribution $\tilde{n}_c(k_x,k_y)$ at $t=40~\mathrm{ps}$ in the $k_x$-$k_y$ plane. (c) Time evolution of the centre of mass coordinates $(\langle x\rangle(t),\langle y \rangle(t))$ of the condensate. Here, $\eta=\{x,y\}$. (d) Trajectory of centre of mass position $(\langle x\rangle(t),\langle y \rangle(t))$ of the condensate. (e) Current dynamics $J_x(t)$ and $J_y(t)$. Here, $J_0=\sqrt{\hbar\gamma_c/m_x}.$ Other parameters are the same as in Fig. \ref{fig:1}, except $\sigma=2.36~\mu \mathrm{m}$ and $m_y/m_x=4$.}
    \label{fig:8}
\end{figure}

For larger pump widths $\sigma$, as we increase the mass ratio $m_y /m_x$, a similar morphing of the condensate width $w_y$ occurs along with the transformation from a circular phase profile to an elliptical phase profile. However, this behavior persists only up to a critical mass ratio $m_c$. Fig. \ref{fig:6}(b) illustrates the change in the critical mass ratio $m_c$ with the pump width $\sigma$. Above this threshold value $m_c$ for different beam widths, the system enters the modulationally unstable regime, and pattern formation is observed in the time evolution of the condensate.
To illustrate this behavior, we exemplify for $m_y/m_x=4$ and $\sigma=2.36~\mu\mathrm{m}$ in Fig~\ref{fig:7}. Starting from a random noise as an initial state, the system first evolves into a Gaussian-shaped condensate [see Fig. \ref{fig:7}(a1)]. As time progresses, the condensate becomes increasingly squeezed along the $y$-direction, and around $t\sim 25~\mathrm{ps}$ the stripe patterns emerge together with the rapid fluctuation in the condensate population $N_c$ [see Fig. \ref{fig:7} and Fig. \ref{fig:8}(a)]. The stripe pattern repeatedly appears and disappears during the dynamics, reflecting a dynamically unstable non-equilibrium state [see Fig. \ref{fig:7}]. Along with this, the formation of stripe pattern along the $y$-direction indicates the emergence of directional modulational instability.
The origin of this instability can be interpreted from the Bogoliubov analysis of the homogeneously pumped system. For $m_y/m_x=4$, the most unstable wavelength in the $y$-direction is $\lambda_y^m=2.60~\mu \mathrm{m}$, which becomes comparable to the beam width $\sigma=2.36~\mu \mathrm{m}$ . Indeed, the different masses along the $x$- and $ y$-directions break the translational symmetry of the system and excite the system to higher polariton modes, thereby inducing stripe formation along the $y$-direction. In Fig. \ref{fig:8}(b), we plot the momentum space density $\tilde{n}_c(k_x=0,k_y)$ at $t=6~\mathrm{ps}$ and $t=40~\mathrm{ps}$. The figure highlights that at $t=6~\mathrm{ps}$ the momentum distribution exhibits a single peak centered at $k_y=0~\mu\mathrm{m}^{-1}$. However, at $t=40~\mathrm{ps}$, in addition to the density peak at $k_y=0~\mu\mathrm{m}^{-1}$, two additional side peaks are clearly visible at $k_y=\pm2.60 ~\mu\mathrm{m}^{-1}$. This result agrees with our calculations of the most unstable wave vector $k_y^{m}$ from linear stability analysis for homogeneous pumping [see Fig. \ref{fig:1}(a)]. These symmetric side peaks in the momentum axis $k_y$ essentially indicate a stripe density modulation in real space along the $y$-direction with a wavelength $\lambda_y=2.42~\mu \rm{m}$. In addition, the pronounced temporal fluctuations in the number of condensate particles $N_c$ [see Fig. \ref{fig:8}(a)] confirm the unstable nature of the condensate.

Along with the emergence of the stripe pattern, the condensate oscillates in the $y$-direction. To characterize the motion of the condensate, we plot the variation of the center of mass, $\langle\mathbf{r}_{\mathrm{cm}}\rangle=(\langle x\rangle(t),\langle y \rangle(t))$, of the condensate along the $x$- and $y$-axes with time in Fig. \ref{fig:8}(c). The time evolution of the centre of mass position  is calculated using the following formulas:
\begin{equation}
\begin{split}
    \langle x \rangle(t)=\frac{\int d\mathbf{r}~x|\psi_c(\mathbf{r},t)|^2}{\int d\mathbf{r}~|\psi_c(\mathbf{r},t)|^2} \\
    \langle y \rangle(t)=\frac{\int d\mathbf{r}~y|\psi_c(\mathbf{r},t)|^2}{\int d\mathbf{r}~|\psi_c(\mathbf{r},t)|^2}.
    \label{com}
    \end{split}
\end{equation}
 As shown in Fig. \ref{fig:8}(c), $\langle x \rangle$ is always zero. Whereas, $\langle y \rangle$ begins to oscillate after $t\sim 25~ \mathrm{ps}$, coinciding with the onset of stripe formation. The motion of the condensate favors along the $y$-direction due to mass anisotropy ($m_y>m_x$). The trajectory of the condensate center of mass can also be viewed as a Lissajous figure in a parametric plot in the $(\langle x\rangle(t),\langle y \rangle(t))$ plane, as presented in Fig. \ref{fig:8}(d).

\begin{figure*}[t]
	\centering
	\includegraphics[width=\textwidth]{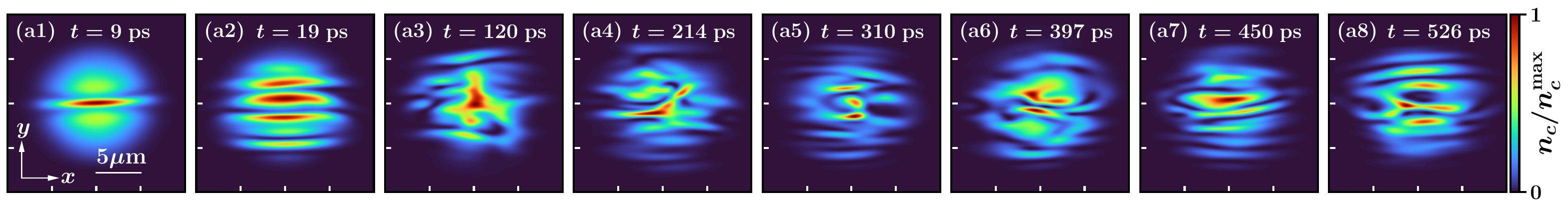}
	\caption{Pattern formation due to modulational instability. (a1)-(a8) show the condensate density $n_c(\mathbf{r},t)=|\psi_c(\mathbf{r},t)|^2$ at different time instants for mass ratio $m_y/m_x=4$ and beam width $\sigma=4.71~\mu\mathrm{m}$. Other parameters are the same as in Fig. \ref{fig:1}}.
    \label{fig:9}
\end{figure*}
Anisotropy in the mass induces oscillation in the condensate dynamics, which can be regarded as self-oscillation, because the translational symmetry of the system is spontaneously broken, and oscillation of the center-of-mass persists without any external oscillating force. The imbalance between the condensate and reservoir density profiles generates a feedback mechanism that sustains the oscillatory motion. In particular, the high-density region of the reservoir tends to drive the stripe pattern toward the direction where the condensate receives a lower particle supply [see Fig. \ref{fig:7}]. Thus, the coupled condensate-reservoir dynamics, together with the mass anisotropy, provide the mechanism for the emergence and persistence of self-oscillations. Due to this self-oscillation, the current dynamics in the $x$- and $y$-directions, $J_x(t)$ and $J_y(t)$, respectively, exhibit distinct behavior. The current dynamics are calculated using the following form
\begin{equation}
    J_{x(y)}(t)=\frac{\hbar}{m_{x(y)}}\frac{\int \dd{\vb{r}}~ \mathrm{Im}\Big(\psi_c^*(\vb{r},t) \nabla_{x(y)}\psi_c(\vb{r},t)\Big)}{\int \dd{\vb{r}}~\abs{\psi_c(\vb{r},t)}^2}. \label{current}
 \end{equation}
 Fig. \ref{fig:8}(e) shows the time evolution of the currents $J_x(t)$ and $J_y(t)$. While $J_x(t)$ remains zero during the entire evolution, $J_y(t)$ exhibits oscillations after $t\sim 25 ~\mathrm{ps}$, linking with the different flow velocities along the $x$- and $y$-direction.

\begin{figure}[h]
	\centering
	\includegraphics[width=0.48\textwidth]{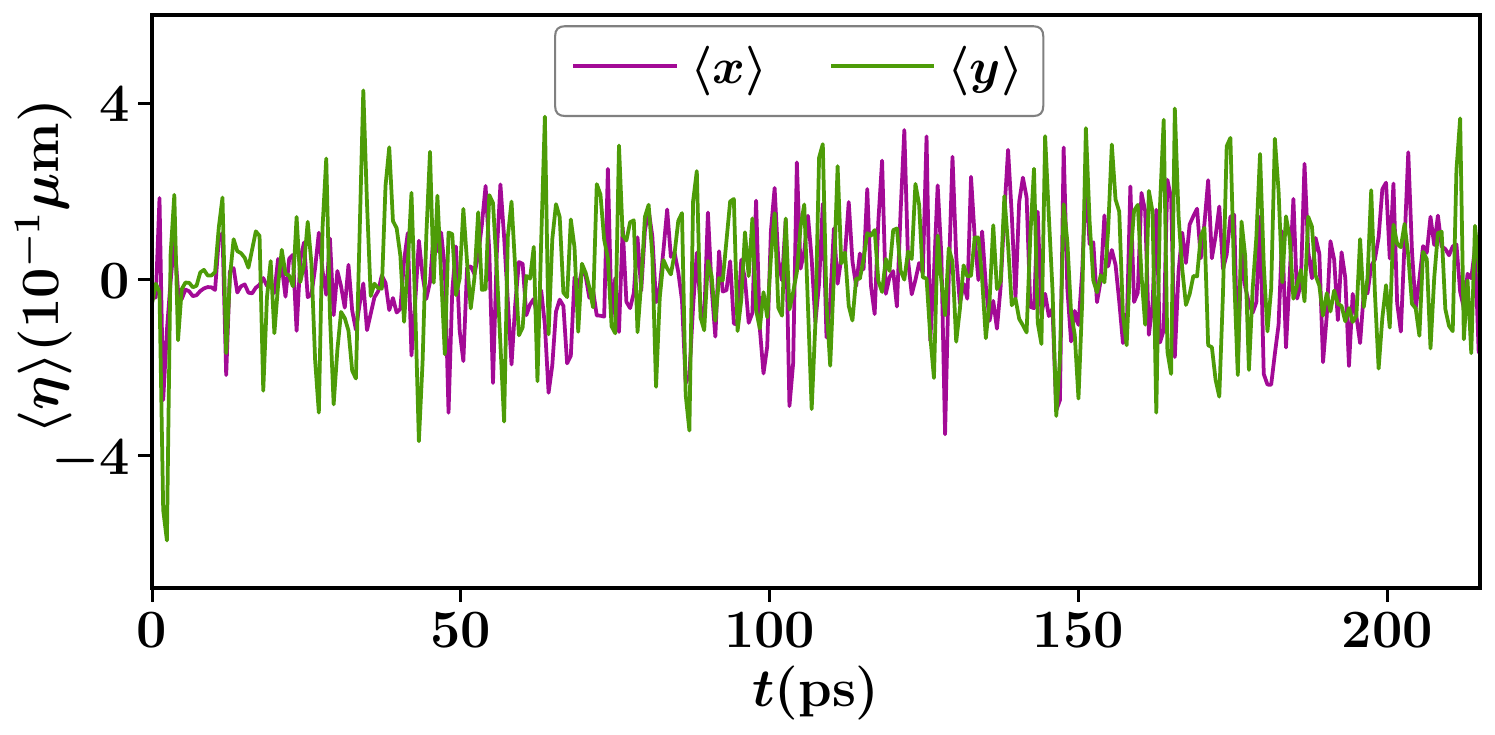}
	\caption{Time evolution of the centre of mass coordinates $(\langle x\rangle(t),\langle y \rangle(t))$ of the condensate. Other parameters are the same as in Fig. \ref{fig:1}, except $\sigma=4.71~\mu \mathrm{m}$ and $m_y/m_x=4$.}
    \label{fig:10}
\end{figure}
So far, we have discussed the effect of mass anisotropy for $\sigma < \lambda_y^m$ and $\sigma \approx \lambda_y^m$. In these regimes, the system exhibits either self-oscillatory ($\sigma \approx \lambda_y^m$) or stable dynamics ($\sigma<\lambda_y^m$), rather than a fully developed modulationally unstable state. It is therefore interesting to explore the influence of mass anisotropy on the regime $\sigma > \lambda_y^m$, where the system is expected to undergo modulational instability. For instance, we have considered $m_y/m_x=4$ and pump width $\sigma=4.71~\mu \mathrm{m}$, which is greater than the most unstable wavelength in the
$y$-direction, i.e, $\lambda_y^m=2.60~\mu \mathrm{m}$. Starting from a random noise as an initial state, the system first evolves into a condensate with a stripe pattern [see Figs. \ref{fig:9}(a1) and \ref{fig:9}(a2)]. Due to mass anisotropy, the stripes initially develop along the $y$-direction. As the system evolves, the stripe orientation becomes dynamically distorted due to modulation instability [see Figs. \ref{fig:9}]. In contrast to the isotropic mass case, where the modulation instability leads to a completely random density distribution [see Figs. \ref{fig:3}(d1) and \ref{fig:3}(d2)], the mass anisotropy introduces a preferred direction, resulting in a more structured evolution of the density pattern, as shown in Figs. \ref{fig:9}. This behavior indicates the emergence of a directional modulational instability. This dynamical evolution is accompanied by persistent oscillations of the condensate center of mass, $\langle x \rangle$ and $\langle y \rangle$, as shown in Fig. \ref{fig:10}, rather than oscillations being confined primarily to the $y$ direction.

\section{Conclusions}
\label{conclusion}
In this study, we examine the impact of mass anisotropy on a non-resonantly pumped polariton condensate. We show that anisotropic effective masses along two orthogonal directions, namely the $x$- and $y$-directions, lead to the emergence of stripe patterns under Gaussian pumping. These patterns arise above a critical mass ratio and are accompanied by fluctuations in the condensate population as well as center-of-mass oscillations, signaling spontaneous breaking of translational symmetry. Notably, both the stripe formation and the oscillatory dynamics of center-of-mass and current preferentially develop along the direction of larger effective mass. On the other hand, for an isotropic mass distribution, a Gaussian pump with a width exceeding the most unstable wavelength of the corresponding isotropic homogeneous system leads to modulational instability. In this regime, the resulting pattern formation, current flow, and center-of-mass dynamics occur randomly without any preferred spatial direction. In contrast, the presence of mass anisotropy can drive the condensate into a modulationally unstable regime even when the Gaussian pump width is smaller than the most unstable wavelength of the isotropic homogeneous system. Consequently, the emergent stripe patterns and collective dynamics become strongly direction-dependent.

To elucidate the underlying mechanism, we perform a linear stability analysis of the homogeneously pumped system. We find that, while mass anisotropy does not modify the instability criteria $P_{\rm{th}}<P<\frac{g_r\gamma_c}{g_c\gamma_r}P_{\rm{th}}$, it significantly alters the dispersion of unstable modes: the most unstable wavelength decreases with increasing mass ratio $m_y/m_x$. This provides a natural length scale for pattern selection. When the pump width becomes comparable to or exceeds this characteristic wavelength, the system enters a directionally modulationally unstable regime, leading to the formation of stripe patterns along the greater mass direction. Our results highlight the crucial role of mass anisotropy in controlling pattern formation and collective dynamics in driven-dissipative exciton-polariton condensates. 

Our work provides a new pathway for exploring nonequilibrium pattern formation in mass-anisotropic polariton condensates and opens up several directions for future research. A natural extension would be to investigate the stability of the soliton structures and to explore the formation of elliptic vortex cores induced by mass anisotropy. It would also be interesting to study how mass anisotropy deforms vortex lattices in mass-anisotropic polariton condensates.

\section*{ACKNOWLEDGMENTS}
We acknowledge the National Supercomputing Mission for
providing computing resources of PARAM Shakti at IIT
Kharagpur. H.S.G. gratefully acknowledges the support from
the Prime Minister’s Research Fellowship (PMRF), India. S.M. appreciates the financial support from the Science and Engineering Research Board (SERB) MATRICS project under Grant No. MTR/2023/000457. S.D. acknowledges support from Grants
No. AFOSR (FA9550-23-1-0034) and ARO (W911NF2210247).

\bibliographystyle{apsrev4-2}
\bibliography{reference.bib}

@article{zhai_2023_electrically,
  title = {Electrically Controlling Vortices in a Neutral Exciton-Polariton Condensate at Room Temperature},
  author = {Zhai, Xiaokun and Ma, Xuekai and Gao, Ying and Xing, Chunzi and Gao, Meini and Dai, Haitao and Wang, Xiao and Pan, Anlian and Schumacher, Stefan and Gao, Tingge},
  journal = {Phys. Rev. Lett.},
  volume = {131},
  issue = {13},
  pages = {136901},
  numpages = {6},
  year = {2023},
  month = {Sep},
  publisher = {American Physical Society},
  doi = {10.1103/PhysRevLett.131.136901},
  url = {https://link.aps.org/doi/10.1103/PhysRevLett.131.136901}
}

@article{wouters_2007_excitations,
  title = {Excitations in a Nonequilibrium Bose-Einstein Condensate of Exciton Polaritons},
  author = {Wouters, Michiel and Carusotto, Iacopo},
  journal = {Phys. Rev. Lett.},
  volume = {99},
  issue = {14},
  pages = {140402},
  numpages = {4},
  year = {2007},
  month = {Oct},
  publisher = {American Physical Society},
  doi = {10.1103/PhysRevLett.99.140402},
  url = {https://link.aps.org/doi/10.1103/PhysRevLett.99.140402}
}

@article{carusotto_2013_quantum,
  title = {Quantum fluids of light},
  author = {Carusotto, Iacopo and Ciuti, Cristiano},
  journal = {Rev. Mod. Phys.},
  volume = {85},
  issue = {1},
  pages = {299--366},
  numpages = {0},
  year = {2013},
  month = {Feb},
  publisher = {American Physical Society},
  doi = {10.1103/RevModPhys.85.299},
  url = {https://link.aps.org/doi/10.1103/RevModPhys.85.299}
}

@article{wouters_2009_stochastic,
  title = {Stochastic classical field model for polariton condensates},
  author = {Wouters, Michiel and Savona, Vincenzo},
  journal = {Phys. Rev. B},
  volume = {79},
  issue = {16},
  pages = {165302},
  numpages = {10},
  year = {2009},
  month = {Apr},
  publisher = {American Physical Society},
  doi = {10.1103/PhysRevB.79.165302},
  url = {https://link.aps.org/doi/10.1103/PhysRevB.79.165302}
}

@article{byrnes_2012_negative,
  title = {Negative Bogoliubov dispersion in exciton-polariton condensates},
  author = {Byrnes, Tim and Horikiri, Tomoyuki and Ishida, Natsuko and Fraser, Michael and Yamamoto, Yoshihisa},
  journal = {Phys. Rev. B},
  volume = {85},
  issue = {7},
  pages = {075130},
  numpages = {5},
  year = {2012},
  month = {Feb},
  publisher = {American Physical Society},
  doi = {10.1103/PhysRevB.85.075130},
  url = {https://link.aps.org/doi/10.1103/PhysRevB.85.075130}
}

@book{pitaevskii_2016_bose,
  title={Bose-Einstein condensation and superfluidity},
  author={Pitaevskii, Lev and Stringari, Sandro},
  volume={164},
  year={2016},
  publisher={Oxford University Press}
}

@book{pethick_2008_bose,
  title={Bose--Einstein condensation in dilute gases},
  author={Pethick, Christopher J and Smith, Henrik},
  year={2008},
  publisher={Cambridge university press}
}

@article{bobrovska_2019_critical,
  title = {Critical dynamics and tree-like spatiotemporal patterns in exciton-polariton condensates},
  author = {Bobrovska, Nataliya and Opala, Andrzej and Mi\ifmmode \mbox{\k{e}}\else \k{e}\fi{}tki, Pawe\l{} and Kulczykowski, Micha\l{} and Szymczak, Piotr and Wouters, Michiel and Matuszewski, Micha\l{}},
  journal = {Phys. Rev. B},
  volume = {99},
  issue = {20},
  pages = {205301},
  numpages = {8},
  year = {2019},
  month = {May},
  publisher = {American Physical Society},
  doi = {10.1103/PhysRevB.99.205301},
  url = {https://link.aps.org/doi/10.1103/PhysRevB.99.205301}
}

@article{saito_2016_selfrotation,
  title = {Self-rotation and synchronization in exciton-polariton condensates},
  author = {Saito, Hiroki and Kanamoto, Rina},
  journal = {Phys. Rev. B},
  volume = {94},
  issue = {16},
  pages = {165306},
  numpages = {6},
  year = {2016},
  month = {Oct},
  publisher = {American Physical Society},
  doi = {10.1103/PhysRevB.94.165306},
  url = {https://link.aps.org/doi/10.1103/PhysRevB.94.165306}
}

@article{bobrovska_2018_dynamical,
  title={Dynamical instability of a nonequilibrium exciton-polariton condensate},
  author={Bobrovska, Nataliya and Matuszewski, Micha{\l} and Daskalakis, Konstantinos S and Maier, Stefan A and K{\'e}na-Cohen, St{\'e}phane},
  journal={ACS Photonics},
  volume={5},
  number={1},
  pages={111--118},
  year={2018},
  publisher={ACS Publications},
  doi={10.1021/acsphotonics.7b00283},
  url={https://pubs.acs.org/doi/10.1021/acsphotonics.7b00283}
}

@article{estrecho_2018_single,
  title={Single-shot condensation of exciton polaritons and the hole burning effect},
  author={Estrecho, E and Gao, Tingge and Bobrovska, Nataliya and Fraser, Michael D and Steger, M and Pfeiffer, L and West, K and Liew, Timothy Chi Hin and Matuszewski, Michal and Snoke, David W and others},
  journal={Nature communications},
  volume={9},
  number={1},
  pages={2944},
  year={2018},
  publisher={Nature Publishing Group UK London},
  doi={10.1038/s41467-018-05349-4},
  url={https://doi.org/10.1038/s41467-018-05349-4}
}

@article{baboux_2018_unstable,
  title={Unstable and stable regimes of polariton condensation},
  author={Baboux, Florent and De Bernardis, Daniele and Goblot, Valentin and Gladilin, VN and Gomez, Carmen and Galopin, E and Le Gratiet, L and Lema{\^\i}tre, A and Sagnes, I and Carusotto, I and others},
  journal={Optica},
  volume={5},
  number={10},
  pages={1163--1170},
  year={2018},
  publisher={Optical Society of America},
  doi={10.1364/OPTICA.5.001163},
  url={https://doi.org/10.1364/OPTICA.5.001163}
}

@article{bobrovska_2014_stability,
  title = {Stability and spatial coherence of nonresonantly pumped exciton-polariton condensates},
  author = {Bobrovska, Nataliya and Ostrovskaya, Elena A. and Matuszewski, Micha\l{}},
  journal = {Phys. Rev. B},
  volume = {90},
  issue = {20},
  pages = {205304},
  numpages = {6},
  year = {2014},
  month = {Nov},
  publisher = {American Physical Society},
  doi = {10.1103/PhysRevB.90.205304},
  url = {https://link.aps.org/doi/10.1103/PhysRevB.90.205304}
}

@article{smirnov_2014_dynamics,
  title = {Dynamics and stability of dark solitons in exciton-polariton condensates},
  author = {Smirnov, Lev A. and Smirnova, Daria A. and Ostrovskaya, Elena A. and Kivshar, Yuri S.},
  journal = {Phys. Rev. B},
  volume = {89},
  issue = {23},
  pages = {235310},
  numpages = {11},
  year = {2014},
  month = {Jun},
  publisher = {American Physical Society},
  doi = {10.1103/PhysRevB.89.235310},
  url = {https://link.aps.org/doi/10.1103/PhysRevB.89.235310}
}

@article{berman_2015_spontaneous,
  title = {Spontaneous formation and nonequilibrium dynamics of a soliton-shaped Bose-Einstein condensate in a trap},
  author = {Berman, Oleg L. and Kezerashvili, Roman Ya. and Kolmakov, German V. and Pomirchi, Leonid M.},
  journal = {Phys. Rev. E},
  volume = {91},
  issue = {6},
  pages = {062901},
  numpages = {12},
  year = {2015},
  month = {Jun},
  publisher = {American Physical Society},
  doi = {10.1103/PhysRevE.91.062901},
  url = {https://link.aps.org/doi/10.1103/PhysRevE.91.062901}
}

@article{weisbuch_1992_observation,
  title = {Observation of the coupled exciton-photon mode splitting in a semiconductor quantum microcavity},
  author = {Weisbuch, C. and Nishioka, M. and Ishikawa, A. and Arakawa, Y.},
  journal = {Phys. Rev. Lett.},
  volume = {69},
  issue = {23},
  pages = {3314--3317},
  numpages = {0},
  year = {1992},
  month = {Dec},
  publisher = {American Physical Society},
  doi = {10.1103/PhysRevLett.69.3314},
  url = {https://link.aps.org/doi/10.1103/PhysRevLett.69.3314}
}

@book{kavokin_2007_microcavities,
  title={Microcavities},
  author={Kavokin, Alexey and Baumberg, Jeremy J and Malpuech, Guillaume and Laussy, Fabrice P},
  year={2007},
  publisher={Oxford University Press}
}

@article{kasprzak_2006_bose,
  title={Bose--Einstein condensation of exciton polaritons},
  author={Kasprzak, Jacek and Richard, Murielle and Kundermann, Stefan and Baas, A and Jeambrun, P and Keeling, Jonathan Mark James and Marchetti, Francesca Maria and Szyma{\'n}ska, MH and Andr{\'e}, R and Staehli, JL a and others},
  journal={Nature},
  volume={443},
  number={7110},
  pages={409--414},
  year={2006},
  publisher={Nature Publishing Group UK London},
  doi={10.1038/nature05131},
  url={https://doi.org/10.1038/nature05131}
}

@article{christopoulos_2007_room,
  title = {Room-Temperature Polariton Lasing in Semiconductor Microcavities},
  author = {Christopoulos, S. and von H\"ogersthal, G. Baldassarri H\"oger and Grundy, A. J. D. and Lagoudakis, P. G. and Kavokin, A. V. and Baumberg, J. J. and Christmann, G. and Butt\'e, R. and Feltin, E. and Carlin, J.-F. and Grandjean, N.},
  journal = {Phys. Rev. Lett.},
  volume = {98},
  issue = {12},
  pages = {126405},
  numpages = {4},
  year = {2007},
  month = {Mar},
  publisher = {American Physical Society},
  doi = {10.1103/PhysRevLett.98.126405},
  url = {https://link.aps.org/doi/10.1103/PhysRevLett.98.126405}
}

@article{kena_2010_room,
  title={Room-temperature polariton lasing in an organic single-crystal microcavity},
  author={K{\'e}na-Cohen, St{\'e}phane and Forrest, SR},
  journal={Nature Photonics},
  volume={4},
  number={6},
  pages={371--375},
  year={2010},
  publisher={Nature Publishing Group UK London},
  doi={10.1038/nphoton.2010.86},
  url={https://doi.org/10.1038/nphoton.2010.86}
}

@article{plumhof_2014_room,
  title={Room-temperature Bose--Einstein condensation of cavity exciton--polaritons in a polymer},
  author={Plumhof, Johannes D and St{\"o}ferle, Thilo and Mai, Lijian and Scherf, Ullrich and Mahrt, Rainer F},
  journal={Nature materials},
  volume={13},
  number={3},
  pages={247--252},
  year={2014},
  publisher={Nature Publishing Group UK London},
  doi={10.1038/nmat3825},
  url={https://doi.org/10.1038/nmat3825}
}

@article{deng_2002_condensation,
  title={Condensation of semiconductor microcavity exciton polaritons},
  author={Deng, Hui and Weihs, Gregor and Santori, Charles and Bloch, Jacqueline and Yamamoto, Yoshihisa},
  journal={Science},
  volume={298},
  number={5591},
  pages={199--202},
  year={2002},
  publisher={American Association for the Advancement of Science},
  doi={10.1126/science.1074464},
  url={https://www.science.org/doi/10.1126/science.1074464}
}

@article{amo_2009_superfluidity,
  title={Superfluidity of polaritons in semiconductor microcavities},
  author={Amo, Alberto and Lefr{\`e}re, J{\'e}r{\^o}me and Pigeon, Simon and Adrados, Claire and Ciuti, Cristiano and Carusotto, Iacopo and Houdr{\'e}, Romuald and Giacobino, Elisabeth and Bramati, Alberto},
  journal={Nature Physics},
  volume={5},
  number={11},
  pages={805--810},
  year={2009},
  publisher={Nature Publishing Group UK London},
  doi={10.1038/nphys1364},
  url={https://doi.org/10.1038/nphys1364}
}

@article{lien_2015_multistability,
  title = {Multistability and condensation of exciton-polaritons below threshold},
  author = {Lien, Jiun-Yi and Chen, Yueh-Nan and Ishida, Natsuko and Chen, Hong-Bin and Hwang, Chi-Chuan and Nori, Franco},
  journal = {Phys. Rev. B},
  volume = {91},
  issue = {2},
  pages = {024511},
  numpages = {9},
  year = {2015},
  month = {Jan},
  publisher = {American Physical Society},
  doi = {10.1103/PhysRevB.91.024511},
  url = {https://link.aps.org/doi/10.1103/PhysRevB.91.024511}
}

@article{ma_2018_vortex,
  title = {Vortex Multistability and Bessel Vortices in Polariton Condensates},
  author = {Ma, Xuekai and Schumacher, Stefan},
  journal = {Phys. Rev. Lett.},
  volume = {121},
  issue = {22},
  pages = {227404},
  numpages = {6},
  year = {2018},
  month = {Nov},
  publisher = {American Physical Society},
  doi = {10.1103/PhysRevLett.121.227404},
  url = {https://link.aps.org/doi/10.1103/PhysRevLett.121.227404}
}

@article{yu_2021_non,
  title = {Non-Hermitian spectrum and multistability in exciton-polariton condensates},
  author = {Yu, Zi-Fa and Xue, Ju-Kui and Zhuang, Lin and Zhao, Jinkui and Liu, Wu-Ming},
  journal = {Phys. Rev. B},
  volume = {104},
  issue = {23},
  pages = {235408},
  numpages = {10},
  year = {2021},
  month = {Dec},
  publisher = {American Physical Society},
  doi = {10.1103/PhysRevB.104.235408},
  url = {https://link.aps.org/doi/10.1103/PhysRevB.104.235408}
}

@article{nardin_2011_hydrodynamic,
  title={Hydrodynamic nucleation of quantized vortex pairs in a polariton quantum fluid},
  author={Nardin, Ga{\"e}l and Grosso, Gabriele and L{\'e}ger, Yoan and Pi\c{e}tka, Barbara and Morier-Genoud, Fran{\c{c}}ois and Deveaud-Pl{\'e}dran, Beno{\^\i}t},
  journal={Nature Physics},
  volume={7},
  number={8},
  pages={635--641},
  year={2011},
  publisher={Nature Publishing Group UK London},
  doi={10.1038/nphys1959},
  url={https://doi.org/10.1038/nphys1959}
}

@article{amo_2009_collective,
  title={Collective fluid dynamics of a polariton condensate in a semiconductor microcavity},
  author={Amo, Alberto and Sanvitto, D and Laussy, FP and Ballarini, D and Valle, E del and Martin, MD and Lemaitre, A and Bloch, J and Krizhanovskii, DN and Skolnick, MS and others},
  journal={Nature},
  volume={457},
  number={7227},
  pages={291--295},
  year={2009},
  publisher={Nature Publishing Group UK London},
  doi={10.1038/nature07640},
  url={https://doi.org/10.1038/nature07640}
}

@article{amo_2011_polariton,
  title={Polariton superfluids reveal quantum hydrodynamic solitons},
  author={Amo, Alberto and Pigeon, S and Sanvitto, D and Sala, VG and Hivet, R and Carusotto, Iacopo and Pisanello, F and Lem{\'e}nager, G and Houdr{\'e}, R and Giacobino, E and others},
  journal={Science},
  volume={332},
  number={6034},
  pages={1167--1170},
  year={2011},
  publisher={American Association for the Advancement of Science},
  doi={10.1126/science.1202307},
  url={https://www.science.org/doi/10.1126/science.1202307}
}

@article{grosso_2011_soliton,
  title = {Soliton Instabilities and Vortex Street Formation in a Polariton Quantum Fluid},
  author = {Grosso, G. and Nardin, G. and Morier-Genoud, F. and L\'eger, Y. and Deveaud-Pl\'edran, B.},
  journal = {Phys. Rev. Lett.},
  volume = {107},
  issue = {24},
  pages = {245301},
  numpages = {5},
  year = {2011},
  month = {Dec},
  publisher = {American Physical Society},
  doi = {10.1103/PhysRevLett.107.245301},
  url = {https://link.aps.org/doi/10.1103/PhysRevLett.107.245301}
}

@article{lagoudakis_2008_quantized,
  title={Quantized vortices in an exciton--polariton condensate},
  author={Lagoudakis, Konstantinos G and Wouters, Michiel and Richard, Maxime and Baas, Augustin and Carusotto, Iacopo and Andr{\'e}, Regis and Dang, Le Si and Deveaud-Pl{\'e}dran, B},
  journal={Nature physics},
  volume={4},
  number={9},
  pages={706--710},
  year={2008},
  publisher={Nature Publishing Group UK London},
  doi={10.1038/nphys1051},
  url={https://doi.org/10.1038/nphys1051}
}

@article{panico_2021_dynamics,
  title = {Dynamics of a Vortex Lattice in an Expanding Polariton Quantum Fluid},
  author = {Panico, Riccardo and Macorini, Guido and Dominici, Lorenzo and Gianfrate, Antonio and Fieramosca, Antonio and De Giorgi, Milena and Gigli, Giuseppe and Sanvitto, Daniele and Lanotte, Alessandra S. and Ballarini, Dario},
  journal = {Phys. Rev. Lett.},
  volume = {127},
  issue = {4},
  pages = {047401},
  numpages = {6},
  year = {2021},
  month = {Jul},
  publisher = {American Physical Society},
  doi = {10.1103/PhysRevLett.127.047401},
  url = {https://link.aps.org/doi/10.1103/PhysRevLett.127.047401}
}

@article{boulier_2015_vortex,
  title={Vortex chain in a resonantly pumped polariton superfluid},
  author={Boulier, T and Ter{\c{c}}as, H and Solnyshkov, DD and Glorieux, Q and Giacobino, E and Malpuech, G and Bramati, A},
  journal={Scientific reports},
  volume={5},
  number={1},
  pages={9230},
  year={2015},
  publisher={Nature Publishing Group UK London},
  doi={10.1038/srep09230},
  url={https://doi.org/10.1038/srep09230}
}

@article{roumpos_2011_single,
  title={Single vortex--antivortex pair in an exciton-polariton condensate},
  author={Roumpos, Georgios and Fraser, Michael D and L{\"o}ffler, Andreas and H{\"o}fling, Sven and Forchel, Alfred and Yamamoto, Yoshihisa},
  journal={Nature Physics},
  volume={7},
  number={2},
  pages={129--133},
  year={2011},
  publisher={Nature Publishing Group UK London},
  doi={10.1038/nphys1841},
  url={https://doi.org/10.1038/nphys1841}
}

@article{sanvitto_2011_all,
  title={All-optical control of the quantum flow of a polariton condensate},
  author={Sanvitto, D and Pigeon, S and Amo, A and Ballarini, D and De Giorgi, M and Carusotto, Iacopo and Hivet, R and Pisanello, F and Sala, VG and Guimaraes, PSS and others},
  journal={Nature photonics},
  volume={5},
  number={10},
  pages={610--614},
  year={2011},
  publisher={Nature Publishing Group UK London},
  doi={10.1038/nphoton.2011.211},
  url={https://doi.org/10.1038/nphoton.2011.211}
}

@article{dominici_2018_interactions,
  title={Interactions and scattering of quantum vortices in a polariton fluid},
  author={Dominici, Lorenzo and Carretero-Gonz{\'a}lez, Ricardo and Gianfrate, Antonio and Cuevas-Maraver, Jes{\'u}s and Rodrigues, Augusto S and Frantzeskakis, Dimitri J and Lerario, Giovanni and Ballarini, Dario and De Giorgi, Milena and Gigli, Giuseppe and others},
  journal={Nature communications},
  volume={9},
  number={1},
  pages={1467},
  year={2018},
  publisher={Nature Publishing Group UK London},
  doi={10.1038/s41467-018-03736-5},
  url={https://doi.org/10.1038/s41467-018-03736-5}
}

@article{gnusov_2023_quantum,
  title={Quantum vortex formation in the “rotating bucket” experiment with polariton condensates},
  author={Gnusov, Ivan and Harrison, Stella and Alyatkin, Sergey and Sitnik, Kirill and T{\"o}pfer, Julian and Sigurdsson, Helgi and Lagoudakis, Pavlos},
  journal={Science advances},
  volume={9},
  number={4},
  pages={eadd1299},
  year={2023},
  publisher={American Association for the Advancement of Science},
  doi={10.1126/sciadv.add1299},
  url={https://www.science.org/doi/10.1126/sciadv.add1299}
}

@article{gnusov_2024_vortex,
  title = {Vortex clusters in a stirred polariton condensate},
  author = {Gnusov, I. and Harrison, S. and Alyatkin, S. and Sitnik, K. and Sigur\dh{}sson, H. and Lagoudakis, P. G.},
  journal = {Phys. Rev. B},
  volume = {109},
  issue = {10},
  pages = {104503},
  numpages = {7},
  year = {2024},
  month = {Mar},
  publisher = {American Physical Society},
  doi = {10.1103/PhysRevB.109.104503},
  url = {https://link.aps.org/doi/10.1103/PhysRevB.109.104503}
}

@article{keeling_2008_spontaneous,
  title = {Spontaneous Rotating Vortex Lattices in a Pumped Decaying Condensate},
  author = {Keeling, Jonathan and Berloff, Natalia G.},
  journal = {Phys. Rev. Lett.},
  volume = {100},
  issue = {25},
  pages = {250401},
  numpages = {4},
  year = {2008},
  month = {Jun},
  publisher = {American Physical Society},
  doi = {10.1103/PhysRevLett.100.250401},
  url = {https://link.aps.org/doi/10.1103/PhysRevLett.100.250401}
}

@article{borgh_2012_robustness,
  title = {Robustness and observability of rotating vortex lattices in an exciton-polariton condensate},
  author = {Borgh, Magnus O. and Franchetti, Guido and Keeling, Jonathan and Berloff, Natalia G.},
  journal = {Phys. Rev. B},
  volume = {86},
  issue = {3},
  pages = {035307},
  numpages = {11},
  year = {2012},
  month = {Jul},
  publisher = {American Physical Society},
  doi = {10.1103/PhysRevB.86.035307},
  url = {https://link.aps.org/doi/10.1103/PhysRevB.86.035307}
}

@article{sun_2019_emergence,
  title = {Emergence and stability of spontaneous vortex lattices in exciton-polariton condensates},
  author = {Sun, F. X. and Niu, Z. X. and Gong, Q. H. and He, Q. Y. and Zhang, W.},
  journal = {Phys. Rev. B},
  volume = {100},
  issue = {1},
  pages = {014517},
  numpages = {7},
  year = {2019},
  month = {Jul},
  publisher = {American Physical Society},
  doi = {10.1103/PhysRevB.100.014517},
  url = {https://link.aps.org/doi/10.1103/PhysRevB.100.014517}
}

@article{kwon_2019_direct,
  title = {Direct Transfer of Light's Orbital Angular Momentum onto a Nonresonantly Excited Polariton Superfluid},
  author = {Kwon, Min-Sik and Oh, Byoung Yong and Gong, Su-Hyun and Kim, Je-Hyung and Kang, Hang Kyu and Kang, Sooseok and Song, Jin Dong and Choi, Hyoungsoon and Cho, Yong-Hoon},
  journal = {Phys. Rev. Lett.},
  volume = {122},
  issue = {4},
  pages = {045302},
  numpages = {7},
  year = {2019},
  month = {Jan},
  publisher = {American Physical Society},
  doi = {10.1103/PhysRevLett.122.045302},
  url = {https://link.aps.org/doi/10.1103/PhysRevLett.122.045302}
}

@article{yulin_2023_vorticity,
  title = {Vorticity of polariton condensates in rotating traps},
  author = {Yulin, A. V. and Shelykh, I. A. and Sedov, E. S. and Kavokin, A. V.},
  journal = {Phys. Rev. B},
  volume = {108},
  issue = {15},
  pages = {155301},
  numpages = {15},
  year = {2023},
  month = {Oct},
  publisher = {American Physical Society},
  doi = {10.1103/PhysRevB.108.155301},
  url = {https://link.aps.org/doi/10.1103/PhysRevB.108.155301}
}

@article{gorbach_2010_vortex,
  title = {Vortex Lattices in Coherently Pumped Polariton Microcavities},
  author = {Gorbach, A. V. and Hartley, R. and Skryabin, D. V.},
  journal = {Phys. Rev. Lett.},
  volume = {104},
  issue = {21},
  pages = {213903},
  numpages = {4},
  year = {2010},
  month = {May},
  publisher = {American Physical Society},
  doi = {10.1103/PhysRevLett.104.213903},
  url = {https://link.aps.org/doi/10.1103/PhysRevLett.104.213903}
}

@article{gladilin_2019_noise,
  title = {Noise-induced transition from superfluid to vortex state in two-dimensional nonequilibrium polariton condensates},
  author = {Gladilin, Vladimir N. and Wouters, Michiel},
  journal = {Phys. Rev. B},
  volume = {100},
  issue = {21},
  pages = {214506},
  numpages = {8},
  year = {2019},
  month = {Dec},
  publisher = {American Physical Society},
  doi = {10.1103/PhysRevB.100.214506},
  url = {https://link.aps.org/doi/10.1103/PhysRevB.100.214506}
}

@article{raman_2026_dynamically,
  title={Dynamically Stable Vortices in Exciton-Polariton Condensates Engineered by Repulsive Interactions},
  author={Raman, P and Radha, R and Mishra, Pankaj K and Muruganandam, Paulsamy},
  journal={arXiv preprint arXiv:2603.25143},
  year={2026},
  doi={10.48550/arXiv.2603.25143},
  url={https://doi.org/10.48550/arXiv.2603.25143}
}

@article{fraser_2009_vortex,
  title={Vortex--antivortex pair dynamics in an exciton--polariton condensate},
  author={Fraser, Michael D and Roumpos, Georgios and Yamamoto, Yoshihisa},
  journal={New Journal of Physics},
  volume={11},
  number={11},
  pages={113048},
  year={2009},
  doi={10.1088/1367-2630/11/11/113048},
  url={https://iopscience.iop.org/article/10.1088/1367-2630/11/11/113048}
}

@article{ma_2026_vortices,
  title={Vortices and solitons in polariton superfluids and condensates},
  author={Ma, Xuekai and Solnyshkov, Dmitry and Malpuech, Guillaume and Schumacher, Stefan and Kavokin, Alexey},
  journal={Nature Reviews Physics},
  pages={1--15},
  year={2026},
  publisher={Nature Publishing Group UK London},
  doi={10.1038/s42254-026-00943-8},
  url={https://doi.org/10.1038/s42254-026-00943-8}
}

@article{lagoudakis_2009_observation,
  title={Observation of half-quantum vortices in an exciton-polariton condensate},
  author={Lagoudakis, KG and Ostatnick{\`y}, T and Kavokin, Alexey V and Rubo, Yuri G and Andr{\'e}, R{\'e}gis and Deveaud-Pl{\'e}dran, Benoit},
  journal={science},
  volume={326},
  number={5955},
  pages={974--976},
  year={2009},
  publisher={American Association for the Advancement of Science},
  doi={10.1126/science.1177980},
  url={https://www.science.org/doi/10.1126/science.1177980}
}

@article{saito_2012_benard,
  title = {B\'enard--von K\'arm\'an vortex street in an exciton-polariton superfluid},
  author = {Saito, Hiroki and Aioi, Tomohiko and Kadokura, Tsuyoshi},
  journal = {Phys. Rev. B},
  volume = {86},
  issue = {1},
  pages = {014504},
  numpages = {5},
  year = {2012},
  month = {Jul},
  publisher = {American Physical Society},
  doi = {10.1103/PhysRevB.86.014504},
  url = {https://link.aps.org/doi/10.1103/PhysRevB.86.014504}
}

@article{lagoudakis_2011_probing,
  title = {Probing the Dynamics of Spontaneous Quantum Vortices in Polariton Superfluids},
  author = {Lagoudakis, K. G. and Manni, F. and Pietka, B. and Wouters, M. and Liew, T. C. H. and Savona, V. and Kavokin, A. V. and Andr\'e, R. and Deveaud-Pl\'edran, B.},
  journal = {Phys. Rev. Lett.},
  volume = {106},
  issue = {11},
  pages = {115301},
  numpages = {4},
  year = {2011},
  month = {Mar},
  publisher = {American Physical Society},
  doi = {10.1103/PhysRevLett.106.115301},
  url = {https://link.aps.org/doi/10.1103/PhysRevLett.106.115301}
}

@article{ballarini_2013_all,
  title={All-optical polariton transistor},
  author={Ballarini, Dario and De Giorgi, Milena and Cancellieri, Emiliano and Houdr{\'e}, Romuald and Giacobino, Elisabeth and Cingolani, Roberto and Bramati, Alberto and Gigli, Giuseppe and Sanvitto, Daniele},
  journal={Nature communications},
  volume={4},
  number={1},
  pages={1778},
  year={2013},
  publisher={Nature Publishing Group UK London},
  doi={10.1038/ncomms2734},
  url={https://doi.org/10.1038/ncomms2734}
}

@article{zezyulin_2018_spin,
  title={Spin--orbit coupled polariton condensates in a radially periodic potential: multiring vortices and rotating solitons},
  author={Zezyulin, Dmitry A and Kartashov, Yaroslav V and Skryabin, Dmitry V and Shelykh, Ivan A},
  journal={ACS Photonics},
  volume={5},
  number={9},
  pages={3634--3642},
  year={2018},
  publisher={ACS Publications},
  doi={10.1021/acsphotonics.8b00536},
  url={https://pubs.acs.org/doi/10.1021/acsphotonics.8b00536}
}

@article{xue_2014_creation,
  title = {Creation and Abrupt Decay of a Quasistationary Dark Soliton in a Polariton Condensate},
  author = {Xue, Yan and Matuszewski, Micha\l{}},
  journal = {Phys. Rev. Lett.},
  volume = {112},
  issue = {21},
  pages = {216401},
  numpages = {5},
  year = {2014},
  month = {May},
  publisher = {American Physical Society},
  doi = {10.1103/PhysRevLett.112.216401},
  url = {https://link.aps.org/doi/10.1103/PhysRevLett.112.216401}
}

@article{zhang_2021_generation,
  title={Generation and stability of diversiform nonlinear localized modes in exciton--polariton condensates},
  author={Zhang, Kun and Wen, Wen and Lin, Ji and Li, Hui-jun},
  journal={New Journal of Physics},
  volume={23},
  number={3},
  pages={033011},
  year={2021},
  publisher={IOP Publishing},
  doi={10.1088/1367-2630/abe79f},
  url={https://iopscience.iop.org/article/10.1088/1367-2630/abe79f}
}

@article{maitre_2020_dark,
  title = {Dark-Soliton Molecules in an Exciton-Polariton Superfluid},
  author = {Ma\^{\i}tre, Anne and Lerario, Giovanni and Medeiros, Adri\`a and Claude, Ferdinand and Glorieux, Quentin and Giacobino, Elisabeth and Pigeon, Simon and Bramati, Alberto},
  journal = {Phys. Rev. X},
  volume = {10},
  issue = {4},
  pages = {041028},
  numpages = {9},
  year = {2020},
  month = {Nov},
  publisher = {American Physical Society},
  doi = {10.1103/PhysRevX.10.041028},
  url = {https://link.aps.org/doi/10.1103/PhysRevX.10.041028}
}

@article{hu_2024_dark,
  title = {Dark soliton cloning in exciton-polariton condensates},
  author = {Hu, Junwei and Zhang, Kun and Idrees, Muhammad and Li, Hui-jun and Lin, Ji and Kavokin, Alexey},
  journal = {Phys. Rev. B},
  volume = {110},
  issue = {15},
  pages = {155112},
  numpages = {9},
  year = {2024},
  month = {Oct},
  publisher = {American Physical Society},
  doi = {10.1103/PhysRevB.110.155112},
  url = {https://link.aps.org/doi/10.1103/PhysRevB.110.155112}
}

@article{hu_2025_gaussian,
  title = {Gaussian potential driven dark soliton and their cloning in exciton-polariton condensates},
  author = {Hu, Junwei and Idrees, Muhammad and Zhang, Kun and Li, Hui-jun and Lin, Ji and Kavokin, Alexey},
  journal = {Phys. Rev. B},
  volume = {111},
  issue = {16},
  pages = {165142},
  numpages = {12},
  year = {2025},
  month = {Apr},
  publisher = {American Physical Society},
  doi = {10.1103/PhysRevB.111.165142},
  url = {https://link.aps.org/doi/10.1103/PhysRevB.111.165142}
}

@article{ostrovskaya_2012_dissipative,
  title = {Dissipative solitons and vortices in polariton Bose-Einstein condensates},
  author = {Ostrovskaya, Elena A. and Abdullaev, Jasur and Desyatnikov, Anton S. and Fraser, Michael D. and Kivshar, Yuri S.},
  journal = {Phys. Rev. A},
  volume = {86},
  issue = {1},
  pages = {013636},
  numpages = {7},
  year = {2012},
  month = {Jul},
  publisher = {American Physical Society},
  doi = {10.1103/PhysRevA.86.013636},
  url = {https://link.aps.org/doi/10.1103/PhysRevA.86.013636}
}

@article{hu_2026_self,
  title = {Self-localized solitonlike peak in uniform nonresonantly pumped exciton-polariton condensates},
  author = {Hu, Junwei and Idrees, Muhammad and Zhang, Kun and Lin, Ji and Li, Hui-jun and Kavokin, Alexey},
  journal = {Phys. Rev. B},
  volume = {113},
  issue = {7},
  pages = {075157},
  numpages = {10},
  year = {2026},
  month = {Feb},
  publisher = {American Physical Society},
  doi = {10.1103/kw8l-mh6q},
  url = {https://link.aps.org/doi/10.1103/kw8l-mh6q}
}

@article{egorov_2009_bright,
  title = {Bright Cavity Polariton Solitons},
  author = {Egorov, O. A. and Skryabin, D. V. and Yulin, A. V. and Lederer, F.},
  journal = {Phys. Rev. Lett.},
  volume = {102},
  issue = {15},
  pages = {153904},
  numpages = {4},
  year = {2009},
  month = {Apr},
  publisher = {American Physical Society},
  doi = {10.1103/PhysRevLett.102.153904},
  url = {https://link.aps.org/doi/10.1103/PhysRevLett.102.153904}
}

@article{sich_2012_observation,
  title={Observation of bright polariton solitons in a semiconductor microcavity},
  author={Sich, M and Krizhanovskii, DN and Skolnick, MS and Gorbach, Andriy V and Hartley, Robin and Skryabin, Dmitry V and Cerda-M{\'e}ndez, EA and Biermann, K and Hey, R and Santos, PV},
  journal={Nature photonics},
  volume={6},
  number={1},
  pages={50--55},
  year={2012},
  publisher={Nature Publishing Group UK London},
  doi={10.1038/nphoton.2011.267},
  url={https://doi.org/10.1038/nphoton.2011.267}
}

@article{whittaker_2017_polariton,
  title = {Polariton Pattern Formation and Photon Statistics of the Associated Emission},
  author = {Whittaker, C. E. and Dzurnak, B. and Egorov, O. A. and Buonaiuto, G. and Walker, P. M. and Cancellieri, E. and Whittaker, D. M. and Clarke, E. and Gavrilov, S. S. and Skolnick, M. S. and Krizhanovskii, D. N.},
  journal = {Phys. Rev. X},
  volume = {7},
  issue = {3},
  pages = {031033},
  numpages = {12},
  year = {2017},
  month = {Aug},
  publisher = {American Physical Society},
  doi = {10.1103/PhysRevX.7.031033},
  url = {https://link.aps.org/doi/10.1103/PhysRevX.7.031033}
}

@article{grosso_2012_dynamics,
  title = {Dynamics of dark-soliton formation in a polariton quantum fluid},
  author = {Grosso, G. and Nardin, G. and Morier-Genoud, F. and L\'eger, Y. and Deveaud-Pl\'edran, B.},
  journal = {Phys. Rev. B},
  volume = {86},
  issue = {2},
  pages = {020509(R)},
  numpages = {4},
  year = {2012},
  month = {Jul},
  publisher = {American Physical Society},
  doi = {10.1103/PhysRevB.86.020509},
  url = {https://link.aps.org/doi/10.1103/PhysRevB.86.020509}
}

@article{yulin_2008_dark,
  title = {Dark polariton solitons in semiconductor microcavities},
  author = {Yulin, A. V. and Egorov, O. A. and Lederer, F. and Skryabin, D. V.},
  journal = {Phys. Rev. A},
  volume = {78},
  issue = {6},
  pages = {061801(R)},
  numpages = {4},
  year = {2008},
  month = {Dec},
  publisher = {American Physical Society},
  doi = {10.1103/PhysRevA.78.061801},
  url = {https://link.aps.org/doi/10.1103/PhysRevA.78.061801}
}

@article{kamchatnov_2012_oblique,
  title={Oblique solitons generated by the flow of a polariton condensate past an obstacle},
  author={Kamchatnov, AM and Korneev, SV},
  journal={Journal of Experimental and Theoretical Physics},
  volume={115},
  number={4},
  pages={579--585},
  year={2012},
  publisher={Springer},
  doi={10.1134/S1063776112080080},
  url={https://doi.org/10.1134/S1063776112080080}
}

@article{hu_2025_vortex,
  title = {Vortex molecules in exciton-polariton condensates formed by uniform nonresonant pumping},
  author = {Hu, Junwei and Idrees, Muhammad and Zhang, Kun and Lin, Ji and Li, Hui-jun and Kavokin, Alexey},
  journal = {Phys. Rev. B},
  volume = {111},
  issue = {24},
  pages = {245119},
  numpages = {13},
  year = {2025},
  month = {Jun},
  publisher = {American Physical Society},
  doi = {10.1103/PhysRevB.111.245119},
  url = {https://link.aps.org/doi/10.1103/PhysRevB.111.245119}
}

@article{ferrini_2025_driven,
  title = {Driven-dissipative turbulence in exciton-polariton quantum fluids},
  author = {Ferrini, R. and Koniakhin, S. V.},
  journal = {Phys. Rev. B},
  volume = {112},
  issue = {20},
  pages = {205305},
  numpages = {10},
  year = {2025},
  month = {Nov},
  publisher = {American Physical Society},
  doi = {10.1103/khp5-5l3d},
  url = {https://link.aps.org/doi/10.1103/khp5-5l3d}
}

@article{berloff_2010_turbulence,
  title={Turbulence in exciton-polariton condensates},
  author={Berloff, Natalia G},
  journal={arXiv preprint arXiv:1010.5225},
  year={2010},
  doi={10.48550/arXiv.1010.5225},
  url={https://doi.org/10.48550/arXiv.1010.5225}
}

@article{panico_2023_onset,
  title={Onset of vortex clustering and inverse energy cascade in dissipative quantum fluids},
  author={Panico, R and Comaron, P and Matuszewski, M and Lanotte, AS and Trypogeorgos, D and Gigli, G and Giorgi, M De and Ardizzone, V and Sanvitto, D and Ballarini, D},
  journal={Nature Photonics},
  volume={17},
  number={5},
  pages={451--456},
  year={2023},
  publisher={Nature Publishing Group UK London},
  doi={10.1038/s41566-023-01174-4},
  url={https://doi.org/10.1038/s41566-023-01174-4}
}

@article{comaron_2025_dynamics,
  title = {Dynamics of Onsager vortex clustering in decaying turbulent polariton quantum fluids},
  author = {Comaron, P. and Panico, R. and Ballarini, D. and Matuszewski, M.},
  journal = {Phys. Rev. Res.},
  volume = {7},
  issue = {2},
  pages = {L022006},
  numpages = {7},
  year = {2025},
  month = {Apr},
  publisher = {American Physical Society},
  doi = {10.1103/PhysRevResearch.7.L022006},
  url = {https://link.aps.org/doi/10.1103/PhysRevResearch.7.L022006}
}

@article{koniakhin_2020_2d,
  title={2D quantum turbulence in a polariton quantum fluid},
  author={Koniakhin, SV and Bleu, O and Malpuech, G and Solnyshkov, DD},
  journal={Chaos, Solitons \& Fractals},
  volume={132},
  pages={109574},
  year={2020},
  publisher={Elsevier},
  doi={10.1016/j.chaos.2019.109574},
  url={https://doi.org/10.1016/j.chaos.2019.109574}
}

@article{gao_2015_observation,
  title={Observation of non-Hermitian degeneracies in a chaotic exciton-polariton billiard},
  author={Gao, Tiejun and Estrecho, E and Bliokh, KY and Liew, TCH and Fraser, MD and Brodbeck, Sebastian and Kamp, Martin and Schneider, Christian and H{\"o}fling, Sven and Yamamoto, Y and others},
  journal={Nature},
  volume={526},
  number={7574},
  pages={554--558},
  year={2015},
  publisher={Nature Publishing Group UK London},
  doi={10.1038/nature15522},
  url={https://doi.org/10.1038/nature15522}
}

@article{dhara_2025_zero,
  title = {Zero-threshold $\mathcal{PT}\text{-symmetric}$ polariton-Raman laser},
  author = {Dhara, A. and Das, P. and Chakrabarty, D. and Ghosh, K. and Chaudhuri, A. Roy and Dhara, S.},
  journal = {Phys. Rev. B},
  volume = {111},
  issue = {4},
  pages = {L041408},
  numpages = {8},
  year = {2025},
  month = {Jan},
  publisher = {American Physical Society},
  doi = {10.1103/PhysRevB.111.L041408},
  url = {https://link.aps.org/doi/10.1103/PhysRevB.111.L041408}
}

@article{dagvadorj_2015_nonequilibrium,
  title = {Nonequilibrium Phase Transition in a Two-Dimensional Driven Open Quantum System},
  author = {Dagvadorj, G. and Fellows, J. M. and Matyja\ifmmode \acute{s}\else \'{s}\fi{}kiewicz, S. and Marchetti, F. M. and Carusotto, I. and Szyma\ifmmode \acute{n}\else \'{n}\fi{}ska, M. H.},
  journal = {Phys. Rev. X},
  volume = {5},
  issue = {4},
  pages = {041028},
  numpages = {9},
  year = {2015},
  month = {Nov},
  publisher = {American Physical Society},
  doi = {10.1103/PhysRevX.5.041028},
  url = {https://link.aps.org/doi/10.1103/PhysRevX.5.041028}
}

@article{caputo_2018_topological,
  title={Topological order and thermal equilibrium in polariton condensates},
  author={Caputo, Davide and Ballarini, Dario and Dagvadorj, Galbadrakh and S{\'a}nchez Mu{\~n}oz, Carlos and De Giorgi, Milena and Dominici, Lorenzo and West, Kenneth and Pfeiffer, Loren N and Gigli, Giuseppe and Laussy, Fabrice P and others},
  journal={Nature materials},
  volume={17},
  number={2},
  pages={145--151},
  year={2018},
  publisher={Nature Publishing Group UK London},
  doi={10.1038/nmat5039},
  url={https://doi.org/10.1038/nmat5039}
}

@article{dagvadorj_2023_unconventional,
  title = {Unconventional Berezinskii-Kosterlitz-Thouless Transition in the Multicomponent Polariton System},
  author = {Dagvadorj, G. and Comaron, P. and Szyma\ifmmode \acute{n}\else \'{n}\fi{}ska, M. H.},
  journal = {Phys. Rev. Lett.},
  volume = {130},
  issue = {13},
  pages = {136001},
  numpages = {6},
  year = {2023},
  month = {Mar},
  publisher = {American Physical Society},
  doi = {10.1103/PhysRevLett.130.136001},
  url = {https://link.aps.org/doi/10.1103/PhysRevLett.130.136001}
}

@article{comaron_2025_coherence,
  title={Coherence of a non-equilibrium polariton condensate across the interaction-mediated phase transition},
  author={Comaron, P and Estrecho, E and Wurdack, M and Pieczarka, M and Steger, M and Snoke, DW and West, K and Pfeiffer, LN and Truscott, AG and Matuszewski, M and others},
  journal={Communications Physics},
  volume={8},
  number={1},
  pages={94},
  year={2025},
  publisher={Nature Publishing Group UK London},
  doi={10.1038/s42005-025-01977-7},
  url={https://doi.org/10.1038/s42005-025-01977-7}
}

@article{nigro_2025_supersolidity,
  title = {Supersolidity of Polariton Condensates in Photonic Crystal Waveguides},
  author = {Nigro, Davide and Trypogeorgos, Dimitrios and Gianfrate, Antonio and Sanvitto, Daniele and Carusotto, Iacopo and Gerace, Dario},
  journal = {Phys. Rev. Lett.},
  volume = {134},
  issue = {5},
  pages = {056002},
  numpages = {8},
  year = {2025},
  month = {Feb},
  publisher = {American Physical Society},
  doi = {10.1103/PhysRevLett.134.056002},
  url = {https://link.aps.org/doi/10.1103/PhysRevLett.134.056002}
}

@article{trypogeorgos_2025_emerging,
  title={Emerging supersolidity in photonic-crystal polariton condensates},
  author={Trypogeorgos, Dimitrios and Gianfrate, Antonio and Landini, Manuele and Nigro, Davide and Gerace, Dario and Carusotto, Iacopo and Riminucci, Fabrizio and Baldwin, Kirk W and Pfeiffer, Loren N and Martone, Giovanni I and others},
  journal={Nature},
  volume={639},
  number={8054},
  pages={337--341},
  year={2025},
  publisher={Nature Publishing Group UK London},
  doi={10.1038/s41586-025-08616-9},
  url={https://doi.org/10.1038/s41586-025-08616-9}
}

@article{tian_2025_towards,
  title={Towards Room-Temperature Exciton--Polariton Supersolidity Driven by Guided Optical Parametric Oscillation},
  author={Tian, Lingyu and Gan, Yusong and Shi, Ying and Xu, Luobing and Xu, Huawen and Xiong, Qihua},
  journal={Chinese Physics Letters},
  volume={42},
  number={9},
  pages={090405},
  year={2025},
  publisher={Chinese Physical Society and IOP Publishing Ltd},
  doi={10.1088/0256-307X/42/9/090405},
  url={https://iopscience.iop.org/article/10.1088/0256-307X/42/9/090405/meta}
}

@article{zhai_2025_electrically,
  title={Electrically tunable nonrigid moire exciton polariton supersolids at room temperature},
  author={Zhai, Xiaokun and Xing, Junhui Cao Chunzi and Yang, Xinmiao and Zhang, Xinzheng and Dai, Haitao and Wang, Xiao and Pan, Anlian and Schumacher, Stefan and Kavokin, Alexey and Ma, Xuekai and others},
  journal={arXiv preprint arXiv:2504.11057},
  year={2025},
  doi={10.48550/arXiv.2504.11057},
  url={https://doi.org/10.48550/arXiv.2504.11057}
}

@article{figueiredo_2026_supersolid,
  title = {Supersolid light in a semiconductor microcavity},
  author = {Figueiredo, J. L. and Mendon\ifmmode \mbox{\c{c}}\else \c{c}\fi{}a, J. T. and Ter\ifmmode \mbox{\c{c}}\else \c{c}\fi{}as, H.},
  journal = {Phys. Rev. A},
  volume = {113},
  issue = {3},
  pages = {L031303},
  numpages = {6},
  year = {2026},
  month = {Mar},
  publisher = {American Physical Society},
  doi = {10.1103/mqmz-nwqj},
  url = {https://link.aps.org/doi/10.1103/mqmz-nwqj}
}

@article{chen_2025_tunable,
  title = {Tunable roton-like exciton states via magnetic fields in two-dimensional layered systems},
  author = {Chen, Yingda and Lou, Wen-Kai and Chang, Kai},
  journal = {Phys. Rev. B},
  volume = {112},
  issue = {19},
  pages = {195304},
  numpages = {14},
  year = {2025},
  month = {Nov},
  publisher = {American Physical Society},
  doi = {10.1103/km2j-4qfp},
  url = {https://link.aps.org/doi/10.1103/km2j-4qfp}
}

@article{grudinina_2026_collective,
  title={Collective Excitations and Stability of Nonequilibrium Polariton Supersolids},
  author={Grudinina, A and Cao, J and Kavokin, A and Voronova, N and Nalitov, A},
  journal={arXiv preprint arXiv:2604.21353},
  year={2026},
  doi={10.48550/arXiv.2604.21353},
  url={https://doi.org/10.48550/arXiv.2604.21353}
}

@article{deng_2010_exciton,
  title = {Exciton-polariton Bose-Einstein condensation},
  author = {Deng, Hui and Haug, Hartmut and Yamamoto, Yoshihisa},
  journal = {Rev. Mod. Phys.},
  volume = {82},
  issue = {2},
  pages = {1489--1537},
  numpages = {0},
  year = {2010},
  month = {May},
  publisher = {American Physical Society},
  doi = {10.1103/RevModPhys.82.1489},
  url = {https://link.aps.org/doi/10.1103/RevModPhys.82.1489}
}

@article{saltykova_2025_quantum,
  title = {Quantum hydrodynamics of a polariton fluid: Pure energy relaxation terms},
  author = {Saltykova, D. A. and Yulin, A. V. and Shelykh, I. A.},
  journal = {Phys. Rev. B},
  volume = {112},
  issue = {6},
  pages = {L060506},
  numpages = {6},
  year = {2025},
  month = {Aug},
  publisher = {American Physical Society},
  doi = {10.1103/1m12-wvk8},
  url = {https://link.aps.org/doi/10.1103/1m12-wvk8}
}

@article{alperin_2022_emergence,
  title = {Emergence and Ordering of Polygonal Breathers in Polariton Condensates},
  author = {Alperin, Samuel N. and Berloff, Natalia G.},
  journal = {Phys. Rev. Lett.},
  volume = {129},
  issue = {1},
  pages = {015301},
  numpages = {5},
  year = {2022},
  month = {Jun},
  publisher = {American Physical Society},
  doi = {10.1103/PhysRevLett.129.015301},
  url = {https://link.aps.org/doi/10.1103/PhysRevLett.129.015301}
}

@article{saito_2013_order,
  title = {Order-Disorder Oscillations in Exciton-Polariton Superfluids},
  author = {Saito, Hiroki and Aioi, Tomohiko and Kadokura, Tsuyoshi},
  journal = {Phys. Rev. Lett.},
  volume = {110},
  issue = {2},
  pages = {026401},
  numpages = {5},
  year = {2013},
  month = {Jan},
  publisher = {American Physical Society},
  doi = {10.1103/PhysRevLett.110.026401},
  url = {https://link.aps.org/doi/10.1103/PhysRevLett.110.026401}
}

@article{sanvitto_2006_spatial,
  title = {Spatial structure and stability of the macroscopically occupied polariton state in the microcavity optical parametric oscillator},
  author = {Sanvitto, D. and Krizhanovskii, D. N. and Whittaker, D. M. and Ceccarelli, S. and Skolnick, M. S. and Roberts, J. S.},
  journal = {Phys. Rev. B},
  volume = {73},
  issue = {24},
  pages = {241308(R)},
  numpages = {4},
  year = {2006},
  month = {Jun},
  publisher = {American Physical Society},
  doi = {10.1103/PhysRevB.73.241308},
  url = {https://link.aps.org/doi/10.1103/PhysRevB.73.241308}
}

@article{li_2016_azimuthons,
  title = {Azimuthons and pattern formation in annularly confined exciton-polariton Bose-Einstein condensates},
  author = {Li, Guangyao},
  journal = {Phys. Rev. A},
  volume = {93},
  issue = {1},
  pages = {013837},
  numpages = {9},
  year = {2016},
  month = {Jan},
  publisher = {American Physical Society},
  doi = {10.1103/PhysRevA.93.013837},
  url = {https://link.aps.org/doi/10.1103/PhysRevA.93.013837}
}

\end{document}